\documentclass[twocolumn,pra,aps,superscriptaddress]{revtex4-2}
                
    \usepackage{amsmath,amssymb,amsthm, mathtools}
    \usepackage{physics}
    \usepackage{graphicx}
    \usepackage{float}
    \usepackage{hyperref}
    \usepackage{xcolor}
    \usepackage{amsfonts}
    \usepackage{dsfont}
    \usepackage{bbm}
    \newtheorem{definition}{Definition}
    \newcommand{\C}{\mathbb{C}}
    \newcommand{\R}{\mathbb{R}}
    \newcommand{\Z}{\mathbb{Z}}
    \newcommand{\id}{\mathds{1}}
    \newcommand{\calH}{\mathcal{H}}
    
    \newcommand{\calD}{\mathcal{D}}

    \newcommand{\FS}{\mathrm{FS}}
    \newcommand{\BS}{\mathrm{BS}}
    
    \newcommand{\sep}{\mathrm{sep}}
    \newcommand{\mix}{\mathrm{mix}}
    
    \newcommand{\om}{\omega}
    \newcommand{\comment}[1]{}
    \newcommand{\cplxi}{\mathrm i}
    \newcommand{\End}{\mathrm{End}}

    \usepackage{tikz}
    \newtheorem{theorem}{Theorem}
    \newtheorem{proposition}{Proposition}
    \newcommand{\ghzprojector}[1]{|#1\rangle\!\langle#1|}
\definecolor{phasezero}{HTML}{24587A}
\definecolor{phaseone}{HTML}{C55B3E}
\definecolor{phasetwo}{HTML}{3B8065}
\newcommand{\fibersupportpanel}[1]{%
\begin{tikzpicture}[x=0.91cm,y=0.91cm,font=\small]
  \pgfmathsetlengthmacro{\fibermarkerradius}{sqrt(3)*2.6pt}
  \foreach \b in {0,1,2}
    \node at (\b+0.5,3.18) {$\b$};
  \node at (1.5,3.55) {$b$};
  \foreach \a in {0,1,2}
    \node at (-0.20,2.5-\a) {$\a$};
  \node at (-0.55,1.5) {$a$};
  \foreach \a in {0,1,2}{%
    \foreach \b in {0,1,2}{%
      \draw[black!20,fill=black!2] (\b,2-\a) rectangle ++(1,1);
      \draw[black!12] (\b+0.5,2.70-\a)
        -- (\b+0.25,2.28-\a) -- (\b+0.75,2.28-\a) -- cycle;
      \pgfmathtruncatemacro{\occupiedphase}{mod(\a+\b,3)}
      \foreach \phase/\dx/\dy/\phasecolor in
        {0/0.5/0.70/phasezero,1/0.25/0.28/phaseone,2/0.75/0.28/phasetwo}{%
        \fill[black!20] (\b+\dx,2-\a+\dy) circle[radius=1.6pt];
        \ifnum#1=0
          \ifnum\a=0 \ifnum\b=0
            \fill[\phasecolor] (\b+\dx,2-\a+\dy) circle[radius=\fibermarkerradius];
          \fi\fi
        \else
          \ifnum#1=1 \def\occupiedphase{0}\fi
          \ifnum\phase=\occupiedphase
            \fill[\phasecolor] (\b+\dx,2-\a+\dy) circle[radius=2.6pt];
          \fi
        \fi
      }%
    }%
  }%
\end{tikzpicture}%
}

\begin{document}
                
    \title{Multipartite GHZ diagonal states and their entanglement properties}
    
    \author{Luka Burduli}
    
    \affiliation{School of Science, Constructor University Bremen, Campus Ring 1, 28759 Bremen, Germany}
    
    \author{Johannes Moerland}
    
    \affiliation{School of Science, Constructor University Bremen, Campus Ring 1, 28759 Bremen, Germany}
    
    \author{Peter Schupp}
    
    \affiliation{School of Science, Constructor University Bremen, Campus Ring 1, 28759 Bremen, Germany}
    
    \begin{abstract}
        In this paper, we investigate the family of \emph{GHZ diagonal states}, which are mixed multipartite quantum states that are diagonal in a GHZ type basis.
        We show that the class of GHZ diagonal states is highly symmetric and exhibits rich entanglement properties, making it relevant for applications in multipartite quantum key distribution.
    \end{abstract}
    
    \maketitle
    
    \section{Introduction}
    
    Bell-diagonal states form one of the most symmetric and useful classes of
    bipartite mixed quantum states. Their high degree of symmetry makes many questions
    about separability analytically tractable, and their rich entanglement properties make them particularly useful for applications in quantum computing and quantum key distribution \cite{riedel2021bell,batle2011nonlocality,bouwmeester2000physics,ribordy2000long}. The notion of Bell-diagonal states can be extended to general finite-dimensional bipartite systems \cite{moerland}. In this paper, we investigate the generalization to a family of density operators on an $N$-qu$d$it Hilbert space.\par
    The natural analog of Bell states in the multipartite setting are Greenberger-Horne-Zeilinger (GHZ) states, which were first constructed in $3$- and $4$-partite systems in~\cite{greenberger1990bell}.
    In this paper, we shall analyze the family of \emph{GHZ diagonal} states. These states were motivated by experiments and constructed explicitly in a computational basis in~\cite{uchida2015entangled}. While useful for low $N$, the description of GHZ diagonal states in terms of a separable basis becomes unfeasible in the higher partite setting, which led the authors to mainly consider the tripartite setting.\par
    In our work, we construct the $N$-partite GHZ diagonal states in a manifestly symmetric way, allowing us to characterize some of their $k$-partite entanglement properties.
    More explicitly, we define the GHZ diagonal states as the convex hull of projectors onto a GHZ-type basis of the $N$-partite Hilbert space $\mathcal H^{\otimes N}$. This basis can be generated by the standard GHZ state
    \begin{equation}
        \ket{\phi^\mathbf 0} = \frac1{\sqrt{d}}\sum_{i=0}^{d-1}\ket i \otimes \dots \otimes \ket i
        \label{eq:ghz0}
    \end{equation}
    under an appropriate subgroup of the multipartite Heisenberg-Weyl group. We then construct the non-entangling \emph{depolarizing channel} that projects arbitrary quantum states onto the set of GHZ diagonal states.
    
    We investigate the entanglement properties of the GHZ diagonal states. Recall that an $N$-partite density operator $\rho$ is said to be \emph{fully separable} if it can be decomposed according to
    \begin{equation}
        \rho = \sum_i p_i \rho^{(1)}_i\otimes\dots\otimes\rho^{(N)}_i
        \label{eq:fs_def}
    \end{equation}
    for a probability distribution $(p_i)$ and a collection of local density operators $\rho^{(S)}_i$. 
    Analogously, a density operator is said to be \emph{separable with respect to a $k$-partition} $\pi = (\pi_1,\dots,\pi_k)$ of $\{1,\dots,N\}$ if it can be written as
    \begin{equation}
        \rho = \sum_i p_i \rho_i^{(\pi_1)} \otimes \dots \otimes \rho^{(\pi_k)}_i,
        \label{eq:kpartition_def}
    \end{equation}
    where $\rho^{(\pi_\ell)}_i$ are density operators on $\mathcal H^{\otimes \pi_\ell} =\bigotimes_{i\in\pi_\ell} \mathcal H$. Finally, $\rho$ is said to be $k$-separable if it can be written as a convex combination of density operators that are separable with respect to (possibly all) $k$-partitions of $\{1,\dots,N\}$. Clearly, any $k$-separable state is also $(k-1)$-separable.
    This general entanglement hierarchy is well-known~\cite{murao1998multiparticle, dur2000classification,horodecki2009}.
    However, already in the bipartite case, the so-called \emph{separability problem}, i.e., deciding whether a given quantum state is separable, is known to be strongly NP-hard~\cite{gharibian2008strong}. Much less is known about the $k$-separability problem in the multipartite setting. Using the symmetries of the GHZ diagonal class of quantum states, we provide a partial answer to the nature of their $k$-separability properties in the form of multipartite separability criteria.

        Full separability of three-qubit GHZ-diagonal states already admits a complete
    characterization. G\"uhne derived a necessary criterion that accounts for
    the relations among the anti-diagonal matrix elements and supplied numerical
    evidence for its sufficiency \cite{guhne2011full}. Han and Kye subsequently
    proved sufficiency and obtained an explicit characterization in terms of the
    matrix entries, using entanglement witnesses associated with positive
    bilinear maps \cite{han2017separability}. These results also distinguish full
    separability from positivity under partial transpose across every bipartition.
    The broader purpose of our approach is to
    construct separable families and analyze the $k$-separability hierarchy on
    $(\C^d)^{\otimes N}$ for arbitrary $N$ and $d$; the finite families constructed
    here provide inner approximations to the fully separable region.

    Genuine multipartite entanglement is likewise completely characterized for
    $N$-qubit GHZ-diagonal states: the criterion of G\"uhne and Seevinck is
    equivalent to $p_{\max}>1/2$, where $p_{\max}$ is the largest GHZ-basis
    probability \cite{guhne2010genuine}. Their sufficiency proof constructs
    biseparable decompositions using pairs of complementary computational-basis
    strings. In this qubit family, biseparability also coincides with being a PPT
    mixture, namely a convex mixture of states that are PPT across possibly
    different bipartitions: GHZ depolarization preserves PPT across each cut,
    and a GHZ-diagonal qubit state that is PPT across a given cut is separable
    across that cut \cite{guhne2011full}. The complementary-pair construction
    does not directly establish the analogous sufficiency result for $d>2$,
    where each GHZ basis vector contains $d$ computational-basis terms.
    Our general-dimensional analysis provides partial criteria and explicit
    constructions, including the sufficient GME condition $p_{\max}>1/d$
    (Sec.~\ref{subsec:majorization}), strictly $k$-separable families, and
    reduced PPT block tests. We do not assume that PPT mixtures characterize
    biseparability for $d>2$.
    \par

    Previous work illustrates how highly symmetric families of quantum states support precise
    separability analysis. Chen, Jiang, and Xu derived witness-based criteria for
    four-qubit GHZ-diagonal states that are necessary and sufficient for
    three-separability within a symmetric subfamily
    \cite{chen2018precise}. For multi-qubit GHZ-diagonal states, Han and Kye
    identified polytope structures associated with biseparability and separability
    across every bipartition; in particular, the biseparable states form a
    hypersimplex inside the GHZ simplex \cite{han2021polytope}. These qubit results
    provide a geometric reference for the present study of arbitrary local
    dimension, without implying that all higher-dimensional separability regions
    are polytopes.
    
    Apart from their theoretical use in quantum key distribution protocols~\cite{epping2017multi}, GHZ entanglement is already a resource for experimental quantum
    communication. Proietti \emph{et al.}\ demonstrated quantum conference key
    agreement among four users by distributing four-photon GHZ states over optical
    fibers and extracting a shared secret key \cite{proietti2021conference}.
    Generating and maintaining multipartite entanglement across imperfect photonic
    links and quantum memories remains an active quantum-network problem. A recent
    preprint by Goetting \emph{et al.}\ reports heralded GHZ entanglement among
    three remote atomic qubits connected by photonic interconnects
    \cite{goetting2026tripartite}.
    Higher-dimensional multipartite entanglement has also been demonstrated
    experimentally: Cervera-Lierta \emph{et al.}\ generated a
    three-qutrit GHZ state in a superconducting processor
    \cite{cervera2022qutrits}. These developments motivate the study of noisy
    multipartite states beyond the qubit setting.
    
    Restricting to states diagonal in a fixed GHZ basis has both
    computational and operational value. Such a state is specified by $d^N-1$
    independent probabilities, compared to $d^{2N}-1$ real parameters for a
    general density operator. Through the depolarizing channel of
    Sec.~\ref{subsec:channel}, criteria for GHZ-diagonal states also certify
    entanglement of arbitrary input states.

    
    
    The paper is structured as follows. Section~\ref{sec:GHZ} introduces the GHZ
    basis and the GHZ diagonal states, their Fourier description in the
    Heisenberg--Weyl basis, the depolarizing channel, and their symmetries.
    Section~\ref{sec:analytic} derives separability criteria for GHZ diagonal states
    and shows how the depolarizing channel extends them to arbitrary states.
    Section~\ref{sec:ent-prop} constructs explicit fully separable, biseparable, and
    strictly $k$-separable states inside the GHZ simplex, including a
    full-dimensional FS simplex for arbitrary $N$ and $d$, which we make explicit
    for three qutrits. Section~\ref{sec:Conclusion} concludes.

    \label{sec:intro}
    
    \section{GHZ Diagonal States}
    \label{sec:GHZ}
    In this section, we develop the notion of GHZ diagonal density operators. Throughout this paper, we consider an $N$-qu$d$it quantum system. We denote by $\calH\cong\C^d$ the local Hilbert space, and equip it with a computational basis $\{\ket{i}\}_{i=0}^{d-1}\subset \calH$. The composite Hilbert space is $\calH^{\otimes N}$, and the computational basis states will be denoted by
    \begin{equation}
        \ket{i_1}\otimes\dots\otimes\ket{i_N} \eqqcolon\ket{i_1,\dots,i_N}.
    \end{equation}
    \subsection{The Heisenberg--Weyl Basis}
    \label{subsec:hw}
    We denote by $X$ and $Z$ the local unitary \emph{shift} and \emph{clock operator}, respectively, determined uniquely by their action on the computational basis vectors:
    \begin{align}
             X, Z:\,&\mathcal H \to \mathcal H,\\
        &X\ket{i} = \ket{i \oplus 1 },\quad
        Z\ket{i} = \omega^{i}\ket{i}.
    \end{align}
    Here (and for the remainder of this paper), $\omega = \exp (2\pi\cplxi / d) $ denotes the principal root of unity, and $\oplus$ is addition modulo $d$.
    We obtain the local \emph{Heisenberg-Weyl operators}  on $\mathcal H$ by
    \begin{equation}
        \mathcal B  = \left\{X^{\mu} Z^{\nu} \right\}_{\mu,\nu = 0}^{d-1}.
    \end{equation}
    They span the entire algebra $\End(\mathcal H)$ of linear operators on $\mathcal H$.
    The Heisenberg-Weyl operators obey the commutation relations
    \begin{equation}\label{eq:hw-comm}
        Z^\nu X^\mu = \omega^{\mu\cdot\nu}X^\mu Z^\nu.
    \end{equation}
    For convenience, we introduce the shorthand notation
        \begin{equation}
        B_i \coloneqq X^{\lfloor i/d \rfloor}Z^{i},
    \end{equation}
    and identify $i$ with $(\mu,\nu)$ via
    \begin{equation}
        i =d\cdot\mu + \nu \leftrightarrow (\mu,\nu) = (\lfloor i / d \rfloor , i \textrm{ mod } d).
    \end{equation}
    The Heisenberg-Weyl operators are orthonormal with respect to the Hilbert-Schmidt inner product, that is,
    \begin{equation}
        \left\langle B_i, B_j\right\rangle \coloneqq
        \frac{1}{d}\tr\left(\left(B_i\right) ^\dagger B_j \right) =
        \delta_{ij}.
    \end{equation}
    We construct a basis of operators on the composite Hilbert space $\mathcal H^{\otimes N}$ by taking tensor products of the local Heisenberg-Weyl operators, that is,
    \begin{equation}
        \mathcal B^{\otimes N} = \{B_{i_1}\otimes\dots\otimes B_{i_N}\}_{i_1,\dots,i_N = 0}^{d^2-1}.
    \end{equation}
    Together with $\omega^\mu$, $B^{\otimes N}$ constitutes a finite group, which we call \emph{$N$-partite Heisenberg-Weyl group} and denote by $\operatorname{HW}(\mathcal H^{\otimes N})$.\par
    The \emph{correlation hypermatrix} $C(\rho) = (c_{i_1,\dots,i_N})$ of a density operator $\rho$ on $\mathcal H^{\otimes N}$ is defined by
    \begin{equation}\label{eq:corr}
        c_{i_1,\dots,i_N} \coloneqq \trace \left(\left( B_{i_1}^\dagger \otimes\dots\otimes B_{i_N}^\dagger\right) \rho \right),
    \end{equation}
    such that
    \begin{align}
        \rho &=\frac{1}{d^N}\sum_{i_1,\dots,i_N = 0}^{d^2-1} c_{i_1,\dots,i_N}B_{i_1}\otimes\dots\otimes B_{i_N}\\
        &=\frac{1}{d^N}\sum_{\mu_1,\nu_1,\dots,\mu_N,\nu_1 = 0}^{d-1} c_{\mu_1,\nu_1;\dots;\mu_N,\nu_1}\\
        &\qquad \times X^{\mu_1}Z^{\nu_1}\otimes\dots\otimes X^{\mu_N}Z^{\nu_N}\label{eq:corr-hw}
    \end{align}
    For $N=1$, this reproduces the Bloch vector in the basis $\mathcal B$, and for $N=2$, the usual notion of the correlation matrix is recovered.
    \subsection{The Class of GHZ Diagonal States}
    \label{subsec:GHZ_def}
    We now construct a maximally GHZ-like entangled basis of the composite Hilbert space $\mathcal H^{\otimes N}$. To do so, we employ the GHZ state $\ket{\phi^\mathbf 0}\in\calH^{\otimes N}$ of
    Eq.~\eqref{eq:ghz0}.
            For each multi-index $\Lambda = (\lambda_1, \ldots, \lambda_N) \in (\Z_d)^N$,
                define the local unitary operator
                \begin{equation}
                  U_\Lambda \coloneqq Z^{\lambda_1} \otimes X^{\lambda_2} \otimes \cdots \otimes X^{\lambda_N}.
                \end{equation}
                The set of these operators forms an abelian group, henceforth denoted $\mathcal U$.
                \begin{definition}[GHZ basis]
                For $\Lambda \in (\Z_d)^N$, we define the \emph{GHZ basis state}
                \begin{equation}
                  |\phi^\Lambda\rangle \coloneqq U_\Lambda |\phi^\mathbf 0\rangle
                  = \frac{1}{\sqrt{d}} \sum_{i=0}^{d-1}
                    \om^{\lambda_1 i}\,
                    |i,\, i\oplus\lambda_2,\, \ldots,\, i\oplus\lambda_N\rangle,
                  \label{eq:GHZ_state}
                \end{equation}
                \end{definition}
                Using the commutation relation~\eqref{eq:hw-comm}, a direct calculation shows that the family $\{|\phi^\Lambda\rangle\}_{\Lambda \in (\Z_d)^N}$ is an
                orthonormal basis of the composite Hilbert space $\calH^{\otimes N}$.
                We have now gathered all the required preliminaries to define our main object of study:
                \begin{definition}[GHZ-diagonal state]
                A \emph{GHZ-diagonal state} is a density operator $\rho$ on the composite Hilbert space $\calH^{\otimes N}$ which is diagonal in the GHZ basis, i.e.,
                \begin{equation}
                  \rho = \sum_{\Lambda \in (\Z_d)^N} p_\Lambda\,
                         |\phi^\Lambda\rangle\langle\phi^\Lambda|, \qquad
                  p_\Lambda \ge 0, \quad \sum_{\Lambda\in (\Z_d)^N} p_\Lambda = 1.
                  \label{eq:GHZ_rho}
                \end{equation}
                \end{definition}
                The simplex $\Delta_\textrm{GHZ}$ of GHZ states may be coordinated by the probability vector $\mathbf{p} = (p_\Lambda)$.
                The vertices of $\Delta_\textrm{GHZ}$ correspond precisely to the pure GHZ basis states.\par
                Whenever we wish to emphasize the dimensions and the number of involved parties, we shall also write $\Delta_{d^N -1}$ for the GHZ simplex.
                We call $\lambda_1$ the \emph{phase label} and
                $\mathbf a=(a_2,\ldots,a_N)\coloneqq(\lambda_2,\ldots,\lambda_N)$ the
                \emph{shift labels} of $\Lambda$, and write $\Lambda=(\lambda_1,\mathbf a)$;
                for $N=3$ we write $\mathbf a=(a,b)$. We also set $a_1\coloneqq 0$, so that the
                $s$-th factor in Eq.~\eqref{eq:GHZ_state} is $\ket{i\oplus a_s}$.
                For fixed $\mathbf a$, the $d$ basis states $\{(\lambda_1,\mathbf a)\}_{\lambda_1\in\Z_d}$
                form the \emph{fiber} of $\mathbf a$, and
                \begin{equation}
                  q_0(\mathbf a)\coloneqq\sum_{\lambda_1\in\Z_d}p_{\lambda_1,\mathbf a}
                  \label{eq:fiber_mass}
                \end{equation}
                is its \emph{fiber mass}. The subscript anticipates the partial Fourier
                transform $q_t$ of Sec.~\ref{subsec:partition}, for which $q_0$ is the case $t=0$.
            \subsection{The Fourier Picture}
            \label{subsec:fourier}
            We now develop an equivalent characterization of the GHZ states in terms of their HW decomposition. It turns out that this alternative description naturally encodes the Fourier transformation of $(p_\Lambda)$.\par
            Let us first compute the correlation hypermatrix of $\ket{\phi^\mathbf 0}\bra{\phi^\mathbf 0}$ as prescribed in eq.~\eqref{eq:corr-hw}. We find
            \begin{equation}              c_{\mu_1,\nu_1;\dots;\mu_N,\nu_N} 
                = \begin{cases}
            1 &\textrm{ if } \mu_1=\dots=\mu_N \textrm{ and } \\&\quad
             \sum_{S=1}^N \nu_S \equiv 0 \mod d, \\
            0 &\textrm{ else}.
            \end{cases}
            \end{equation}
            For each multi-index $M = (\mu, \nu_1, \ldots, \nu_{N-1}) \in (\Z_d)^N$, define the tensor-product unitary
            \begin{equation}
            T_M \coloneqq X^\mu Z^{\nu_1} \otimes X^\mu Z^{\nu_2} \otimes \cdots
                 \otimes X^\mu Z^{-\sum_{s=1}^{N-1}\nu_s}.
          \label{eq:TM_def}
            \end{equation}
            The family $\mathcal{T} = \{T_M : M \in (\Z_d)^N\}$ is a maximal abelian subgroup of $\operatorname{HW}(\mathcal H^{\otimes N})$.
            Note that by construction, it holds
            \begin{equation}
                \ket{\phi^\mathbf 0}\bra{\phi^\mathbf 0} = \frac1{d^N}\sum_{M\in(\mathbb Z_d)^N}T_M.
            \end{equation}  
            We now compute the HW expansion of a general GHZ state. To that end, we use the Heisenberg-Weyl commutation relations~\eqref{eq:hw-comm} and obtain
        \begin{equation}\label{eq:t-u-comm}
            T_M U_\Lambda = \om^{(M,\Lambda)} U_\Lambda T_M,
        \end{equation}
        where the bilinear pairing $(M,\Lambda) \in \Z_d$ is given by
        \begin{equation}
          (M,\Lambda) = -\mu\lambda_1 + \sum_{s=2}^{N-1}\nu_s\lambda_s
                       - \Bigl(\sum_{s=1}^{N-1}\nu_s\Bigr)\lambda_N.
          \label{eq:ML_pairing}
        \end{equation}
        It follows that
                \begin{align}
            \rho &= \sum_\Lambda p_\Lambda\ket{\phi^\Lambda}\bra{\phi^\Lambda} = \frac1{d^N}\sum_\Lambda p_\Lambda U_\Lambda\left(\sum_M T_M\right)U_\Lambda^\dagger\\
            &=\frac1{d^N}\sum_{M} \left(\sum_\Lambda p_\Lambda\om^{-(M,\Lambda)}\right) T_M = \frac{1}{d^N}\sum_Mf_M T_M,
        \end{align}
        
        where
                \begin{equation}
            f_M = \sum_\Lambda p_\Lambda \omega^{-(M,\Lambda)}
            \label{eq:ghz_fourier_coefficients}
        \end{equation}
        
        is the discrete Fourier transform of the probability density $(p_\Lambda)$, and
        the inverse Fourier transform is given by
        \begin{equation}
            p_\Lambda= \frac1{d^N}\sum_M \omega^{(M,\Lambda)}f_M.
        \end{equation}
        Thus, the HW decomposition of the GHZ states accumulates the Fourier transform $(f_M)$ of the probability density $(p_\Lambda)$ as correlation hypermatrix. 
        Apart from the Fourier picture in the HW basis, the GHZ states satisfy pleasant stabilizer properties with respect to the $T_M$ operators. To illustrate this, let us first compute the action of the $T_M$ operators on the GHZ basis. A straight-forward computation reveals that
        \begin{equation}
            T_M|\phi^\mathbf 0\rangle = \ket{\phi^\mathbf 0}
        \end{equation}
        for all $M$. To compute $T_M\ket{\phi^\Lambda}$, we use the commutation relations~\eqref{eq:t-u-comm} between $T_M$ and $U_\Lambda$, yielding
        \begin{equation}
          T_M\ket{\phi^\Lambda} =      \om^{(M,\Lambda)}\ket{\phi^\Lambda}.
          \label{eq:eigenbasis}
        \end{equation}
        Thus, the GHZ states are simultaneous eigenstates of all $T_M \in \mathcal{T}$. Equivalently, the $T_M$ operators are diagonal in the GHZ basis.\par
        Since the eigenvalues in eq.~\eqref{eq:eigenbasis} have unit modulus, it follows from linearity that
        \begin{equation}
            T_M \rho T_M^\dagger = \rho
            \label{eq:T_invariance}
        \end{equation}
        for any GHZ state $\rho$.
        \subsection{The Depolarizing Channel}
        \label{subsec:channel}
        
        Denote the GHZ projectors by $\Phi_\Lambda \coloneqq \ket{\phi^\Lambda}\!\bra{\phi^\Lambda}$.
        The \emph{depolarizing channel} $\mathcal{E}$ is the completely positive, trace-preserving map that dephases any state $\sigma \in \mathcal{D}(\calH^{\otimes N})$ onto the GHZ-diagonal subspace:
        \begin{equation}
          \mathcal{E}(\sigma)
          = \sum_{\Lambda \in (\Z_d)^N}
            \Phi_\Lambda\,\sigma\,\Phi_\Lambda
          = \sum_{\Lambda} \bra{\phi^\Lambda}\sigma\ket{\phi^\Lambda}\,\Phi_\Lambda.
          \label{eq:channel}
        \end{equation}
        The output probabilities are $p_\Lambda = \bra{\phi^\Lambda}\sigma\ket{\phi^\Lambda}$.
        Equivalently, $\mathcal{E}$ can be viewed as the uniform average over the group $\mathcal{T}$ defined above:
        \begin{equation}
          \mathcal{E}(\sigma) = \frac{1}{d^N}\sum_{M\in(\Z_d)^N} T_M \sigma T_M^\dagger.
          \label{eq:GHZ_depolarization_group}
        \end{equation}
        To see that both forms agree, expand $\sigma = \sum_{\Lambda,\Gamma}\sigma_{\Lambda\Gamma}\ket{\phi^\Lambda}\!\bra{\phi^\Gamma}$ and use the eigenvalue relation~\eqref{eq:eigenbasis}:
        \begin{align}
          &\frac{1}{d^N}\sum_{M\in(\Z_d)^N} T_M\sigma T_M^\dagger\\
          =& \sum_{\Lambda,\Gamma\in(\Z_d)^N}\sigma_{\Lambda\Gamma}
             \!\left(\frac{1}{d^N}\sum_{M\in(\Z_d)^N}\om^{(M,\Lambda-\Gamma)}\right)
             \ket{\phi^\Lambda}\!\bra{\phi^\Gamma} \notag\\
          =& \sum_{\Lambda\in(\Z_d)^N}
             \bra{\phi^\Lambda}\sigma\ket{\phi^\Lambda}\,\Phi_\Lambda,
          \label{eq:GHZ_depolarization_diagonal}
        \end{align}
        where we used that $\sum_{M}\omega^{(M,\Lambda-\Gamma)} = d^N\delta_{\Lambda,\Gamma}$.
        The image of $\mathcal{E}$ is exactly the GHZ-diagonal subspace.
        Since every $T_M$ is a tensor product of local unitaries, $\mathcal E$ maps
        states separable across a partition $\pi$ to states separable across $\pi$.
        In particular, $\mathcal E$ preserves $k$-separability for every $k$ and cannot
        create entanglement. Consequently, if $\mathcal E(\sigma)$ is not
        $k$-separable, neither is $\sigma$. Since $\mathcal E$ may also destroy
        entanglement, this gives a sufficient, not a necessary, detection method.
        \subsection{Symmetries of the GHZ diagonal states}
        The convex set of GHZ diagonal states is invariant under the discrete abelian groups $\mathcal U, \mathcal T$, as well as under the permutation group $S_N$. Let us describe the group actions explicitly.
        \paragraph{Action of $\mathcal U$.}
        By construction of the GHZ basis, we have
        \begin{equation}
            U_\Lambda \ket{\phi^{\Gamma}} = \ket{\phi^{\Gamma+\Lambda}},
        \end{equation}
        where $\Lambda+\Gamma$ denotes component wise addition modulo $d$ of the multiindices. For the GHZ projectors, it follows that
        \begin{equation}
            U_\Lambda \Phi_\Gamma U_\Lambda^\dagger = \Phi_{\Gamma+\Lambda},
        \end{equation}
        which extends to $\Delta_\textrm{GHZ}$ by linearity.
        \paragraph{Action of $\mathcal T$.}
        By Eq.~\eqref{eq:T_invariance}, $\mathcal T$ acts trivially on
        $\Delta_\textrm{GHZ}$; conversely, by Eq.~\eqref{eq:GHZ_depolarization_group},
        $\Delta_\textrm{GHZ}$ is exactly the set of states fixed by $\mathcal T$.
        \paragraph{Action of $S_N$.}
        The symmetric group $S_N$ acts on $\mathcal H^{\otimes N}$ by permuting the local subsystems. We now show that this preserves the GHZ diagonal structure. To that end, note that permutations act trivially on $\ket{\phi^\mathbf 0}$. Let $\sigma$ be a permutation that fixes the first subsystem. Then, it holds
        \begin{equation}
            \sigma\ket{\phi^{\Lambda}} = \ket{\phi^{\sigma(\Lambda)}},
        \end{equation}
        where $\sigma(\Lambda)$ denotes the permutation of the components of $\Lambda\in(\mathbb Z_d)^N$.\par
        Now, let $\sigma_{12}$ be the permutation that exchanges the first and second subsystem. A short calculation shows that
        \begin{equation}
            \sigma_{12}\ket{\phi^{\lambda_1,\dots,\lambda_N}} = \omega^{-\lambda_1\lambda_2}\ket{\phi^{\lambda_1,-\lambda_2,\lambda_3-\lambda_2,\dots,\lambda_N-\lambda_2}}.
        \end{equation}
        Since $S_N$ is generated by $\sigma\in S_n$ that fix the first subsystem together with $\sigma_{12}$, one can check that this extends to a representation of $S_n$ on $\Delta_\textrm{GHZ}$, that is,
        \begin{equation}
            \sigma\left(\sigma'(\Phi_\Lambda)\right) = (\sigma\circ\sigma')(\Phi_\Lambda).
        \end{equation}

        \section{Analytic Separability Criteria}
                \label{sec:analytic}        
        In this section, we make use of the symmetries of the class of GHZ diagonal states to derive analytic multipartite separability criteria.
        \subsection{Separability Hierarchy}
        \label{subsec:hierarchy}
        We use the separability notions defined in the Introduction.
        Fully separable (FS) states are those of the form~\eqref{eq:fs_def}.
        For a partition
        $\pi=\{\pi_1,\ldots,\pi_k\}$ of the parties, let $\calD_\pi$ be the set
        of states separable across that fixed partition,
        Eq.~\eqref{eq:kpartition_def}.  The set of
        $k$-separable states is
        \begin{equation}
          \calD_{k\text{-}\sep}
          \coloneqq
          \operatorname{conv}\!\left(
            \bigcup_{\pi:\,|\pi|=k}\calD_\pi
          \right),
        \end{equation}
        where different terms in the convex mixture may use different
        $k$-partitions.  These sets form the filtration
        \begin{equation}
          \calD_{\FS}=\calD_{N\text{-}\sep}
          \subseteq\calD_{(N-1)\text{-}\sep}
          \subseteq\cdots\subseteq
          \calD_{\BS}
          \subseteq\calD(\calH^{\otimes N}).
        \end{equation}
        Biseparable (BS) states are therefore mixtures of states separable
        across bipartitions, and states outside $\calD_{\BS}$ are genuinely
        multipartite entangled (GME).
        
        A state is \emph{strictly $k$-separable} if it can be expressed as a convex combination of decompositions into $k$ separable blocks, but not into convex combinations of
        $k+1$ blocks.  For $N=3$, strict biseparability simply means BS but
        not FS. The maximally mixed state $\rho_{\mix}=\id/d^N$ is FS and
        lies in the interior of $\calD_{\FS}$.

                \subsection{Schur--Horn Majorization and a Sufficient Criterion for GME}
                \label{subsec:majorization}
    
                For $x,y\in\R^n$, let $x^\downarrow$
                and $y^\downarrow$ denote their components in non-increasing order.
                Then $x$ is \emph{majorized by} $y$, written $x\prec y$, if
                
                \begin{gather}
                  \sum_{i=1}^k x_i^\downarrow \leq\sum_{i=1}^k y_i^\downarrow
                    \quad(k=1,\ldots,n-1),\\
                  \sum_{i=1}^n x_i=\sum_{i=1}^n y_i.
                \end{gather}
                See Ref.~\cite[Secs.~2.1 and~12.5]{bengtsson_zyczkowski_2006}
                for details.
    
                For a
                Hermitian matrix $A$ we may take $\lambda(A)$ to be its
                \emph{ordered eigenvalue vector} and $\delta$ its \emph{ordered
                diagonal} in an arbitrary orthonormal basis.
                The Schur--Horn
                theorem states $\delta\prec\lambda(A)$; 
                Conversely, every vector majorized by
                $\lambda(A)$ occurs as such a diagonal.
                \begin{align}
                \{\operatorname{diag}(UAU^\dagger):U\text{ unitary}\}&=\left\{\delta:\delta\prec\lambda(A)\right\} \notag\\
                  &=\operatorname{conv}\left\{P_\pi\lambda(A):\pi\in S_n\right\}.
                  \label{eq:schur_horn_orbit}
                \end{align}
                Geometrically, the possible
                diagonals form a polytope whose vertices are the permutations of
                $\lambda(A)$, so a change of basis can only spread the eigenvalue
                weights more evenly, e.g., $\max_i\delta_i\leq\max_i\lambda_i(A)$.
                For a GHZ-diagonal state, $\mathbf p=(p_\Lambda)$ is the eigenvalue
                vector in the GHZ basis, and $\max_\Lambda p_\Lambda$ is
                Schur-convex: $x\prec y$ implies $\max_i x_i\leq\max_i y_i$.

                For a pure product state $\ket{\psi}=\ket{\psi_A}\otimes\ket{\psi_B}$ across any bipartition $A|B$, Eq.~\eqref{eq:GHZ_state} gives $\langle\phi^\Lambda|\psi\rangle=\tfrac1{\sqrt d}\sum_i\omega^{-\lambda_1 i}a_ib_i$ with $a_i\coloneqq(\bigotimes_{s\in A}\bra{i\oplus\lambda_s})\ket{\psi_A}$, $b_i\coloneqq(\bigotimes_{s\in B}\bra{i\oplus\lambda_s})\ket{\psi_B}$ orthonormal families in $i$, so by Cauchy--Schwarz
    \begin{equation}
    \big|\langle\phi^\Lambda|\psi\rangle\big|^2\leq\frac1d\Big(\sum_i|a_i|^2\Big)\Big(\sum_i|b_i|^2\Big)\leq\frac1d.
    \label{eq:biseparable_overlap_bound}
    \end{equation}
    Convexity extends this to biseparable mixed states, giving $p_\Lambda\leq1/d$ for all $\Lambda$, which by contraposition yields a sufficient criterion for GME,
    \begin{equation}
      \max_\Lambda p_\Lambda > \frac{1}{d}
      \quad\Longrightarrow\quad
      \rho_P\text{ is GME}.
      \label{eq:sufficient_gme_threshold}
    \end{equation}
    This $1/d$ threshold is used in the channel criterion below.
        For $d=2$, also the converse holds:
    $\rho_P$ is biseparable if and only if $p_{\max}\leq1/2$
    \cite{guhne2010genuine}.
            
                \paragraph{Using the depolarizing channel to detect entanglement.}
        By Sec.~\ref{subsec:channel}, any GHZ-diagonal entanglement criterion
        applied to $\mathcal E(\sigma)$ also certifies entanglement of $\sigma$.
        For the white-noise line
        \begin{equation}
        \rho_\epsilon=(1-\epsilon)\mathcal E(\sigma)+\epsilon\frac{\id}{d^N},
        \end{equation}
        the maximum GHZ probability is
        \begin{equation}
        \max_\Lambda p_\Lambda(\epsilon)
        =
        (1-\epsilon)p_{\Lambda_{\max}}+\frac{\epsilon}{d^N},
        \end{equation}
        so the maximum-probability criterion guarantees GME whenever
        \begin{equation}
        \epsilon<
        \frac{p_{\Lambda_{\max}}-1/d}{p_{\Lambda_{\max}}-1/d^N}
        \qquad (p_{\Lambda_{\max}}>1/d).
        \label{eq:channel_noise_certificate}
        \end{equation}
        As an example, for the three-qutrit coherent input
        \begin{equation}
        \ket{\psi_\theta}
        =
        \cos\theta\,\ket{\phi^{0,0,0}}
        +
        \sin\theta\,\ket{\phi^{0,1,1}},
        \end{equation}
        the channel removes the off-diagonal GHZ coherences. Thus, for
        $\sigma_{\theta,\epsilon}=(1-\epsilon)\ket{\psi_\theta}\!\bra{\psi_\theta}+\epsilon\id/27$,
        one has $p_{\Lambda_{\max}}=c_\theta\coloneqq\max\{\cos^2\theta,\sin^2\theta\}$ and
        \begin{equation}
        p_{\Lambda_{\max}}^{(\theta)}(\epsilon)
        =
        (1-\epsilon)c_\theta+\frac{\epsilon}{27},
        \qquad
        \epsilon_\ast(\theta)
        =
        \frac{c_\theta-\frac13}{c_\theta-\frac1{27}} .
        \end{equation}
        Therefore $0\leq\epsilon<\epsilon_\ast(\theta)$ detects both
        $\mathcal E(\sigma_{\theta,\epsilon})$ and the original non-GHZ-diagonal
        state $\sigma_{\theta,\epsilon}$ as GME. 
        The depolarized GHZ vertex is the
        special case $p_{\Lambda_{\max}}=1$, giving
        $\epsilon<(1-1/d)/(1-1/d^N)$, which is equal to $9/13$ for $N=d=3$.
        
        \begin{figure}[t]
          \centering
          \includegraphics[width=\linewidth]{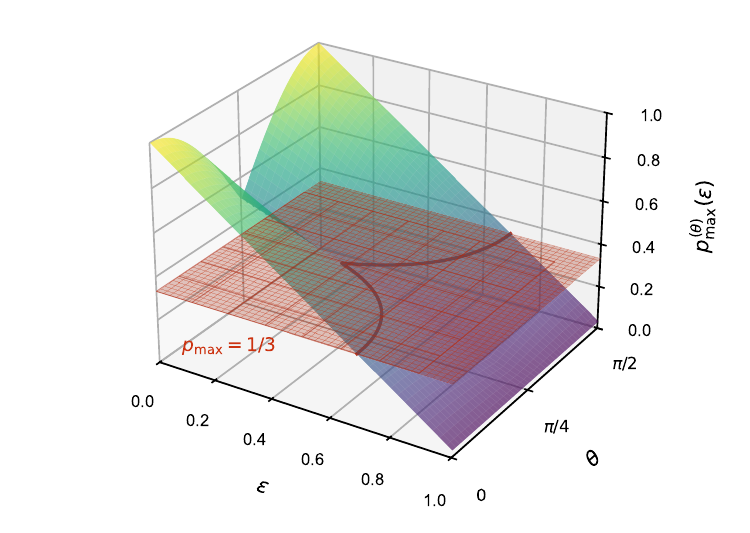}
          \caption{For the three-qutrit coherent-superposition family, the depolarized noisy state satisfies $p_{\max}^{(\theta)}(\epsilon)=(1-\epsilon)c_\theta+\epsilon/27$, where $c_\theta=\max\{\cos^2\theta,\sin^2\theta\}$. The surface is shown over the noise strength $\epsilon$ and angle $\theta$; Horizontal plane denotes the GME threshold $p_{\max}=1/3$, and its intersection with the surface gives the boundary $\epsilon=\epsilon_\ast(\theta)$.}
          \label{fig:channel_detection_pmax}
        \end{figure}
    
        \subsection{Bipartite CCNR and the Multipartite Mixed Fourier Criterion}
                \label{subsec:ccnr}
                The CCNR (realignment) criterion is bipartite and, in a multipartite system,
                may be applied only after choosing a bipartition. For $N=2$, the active support of the Heisenberg--Weyl
                correlation matrix $C(\rho_P)$ is a permutation diagonal and singular values are fourier coefficients, hence
                \begin{equation}
                  \left\|C(\rho_P)\right\|_*
                  =\sum_{M\in\Z_d^2}|f_M|.
                  \label{eq:bipartite_fourier_nuclear}
                \end{equation}
                Consequently, using the fully separable bound
                $\|C(\rho_P)\|_* \le d^{N/2}$ every separable two-party Bell-diagonal state satisfies
                \begin{equation}
                  \sum_{M\in\Z_d^2}|f_M|\leq d
                  \label{eq:ccnr}
                \end{equation}
                (recall that $N=2$ here), and a value exceeding $d$ certifies bipartite entanglement. For $N\geq3$, $C$ is a tensor rather than a matrix, so the equality in
                Eq.~\eqref{eq:bipartite_fourier_nuclear} fails in general.
                
                The valid higher-party replacement is obtained by flattening the correlation
                tensor. For
                $M=(\mu,\nu_1,\nu_2,\ldots,\nu_{N-1})$ define the
                $\ell_{1,2}$ Fourier functional
                \begin{equation}
                  \Phi_1(f)\coloneqq
                  \sum_{\mu,\nu_1\in\Z_d}
                  \left(
                    \sum_{\nu_2,\ldots,\nu_{N-1}\in\Z_d}
                    \left|f_{\mu,\nu_1,\nu_2,\ldots,\nu_{N-1}}\right|^2
                  \right)^{1/2}.
                  \label{eq:mixed_fourier_functional}
                \end{equation}
                The rows of this flattening have disjoint support and are therefore mutually
                orthogonal.  So full separability implies
                \begin{equation}
                  \Phi_1(f)= \left\|\text{Flat}_1C(\rho_P)\right\|_*\leq \left\|C(\rho_P)\right\|_*. \label{eq:mixed_fourier_fs_bound}
                \end{equation}  Thus
                \begin{equation}
                  \Phi_1(f)>d^{N/2}
                  \quad\Longrightarrow\quad
                  \rho_P\text{ is not fully separable}.
                  \label{eq:mixed_fourier_criterion}
                \end{equation}
                
                This nonlinear criterion yields an explicit state-dependent witness.  Let
                $\rho_0$ be a target GHZ-diagonal state with Fourier coefficients $f_M^0$, and
                for each row define
                \begin{equation}
                  R_{\mu,\nu_1}\coloneqq
                  \left(
                    \sum_{\nu_2,\ldots,\nu_{N-1}\in\Z_d}
                    \left|f^0_{\mu,\nu_1,\nu_2,\ldots,\nu_{N-1}}\right|^2
                  \right)^{1/2}.
                \end{equation}
                With the Fourier convention of Sec.~\ref{subsec:fourier}, set
                \begin{equation}
                  A_{\rho_0}\coloneqq\frac12\sum_M
                  \left(\frac{\overline{f_M^0}}{R_{\mu,\nu_1}}T_M^\dagger+ \frac{{f_M^0}}{R_{\mu,\nu_1}}T_M\right).
                  \label{eq:mixed_fourier_supporting_observable}
                \end{equation}
                Then
                 $ \operatorname{Tr}(\rho_0A_{\rho_0})=\Phi_1(f^0),
                $
                Therefore
                \begin{equation}
                  W_{\rho_0}\coloneqq d^{N/2}\id-A_{\rho_0}
                  \label{eq:mixed_fourier_witness}
                \end{equation}
                is an entanglement witness: it is nonnegative on all fully separable
                states. For $N=2$, the
                inner sum in Eq.~\eqref{eq:mixed_fourier_functional} is a singleton, so
                $\Phi_1(f)=\sum_M|f_M|$ and the construction reduces to the ordinary
                bipartite Fourier $\ell_1$ witness.
                
                      \subsection{Partitioned Simplices and PPT Block Criteria}
                \label{subsec:partition}
        
                The global GHZ simplex is only one member of a larger family obtained
                by grouping the parties.  For a partition
                \(\pi=\{\pi_1,\ldots,\pi_\ell\}\), in the notation of
                Sec.~\ref{subsec:hierarchy}, write \(n_a\coloneqq|\pi_a|\).  On a nonsingleton
                block \(\pi_a\), after fixing an ordering of its parties, define the
                block GHZ vectors
                \begin{equation}
                  \ket{\phi_{\pi_a}^{\Lambda_a}}
                  =
                  \frac{1}{\sqrt d}
                  \sum_{j\in\Z_d}
                  \om^{\lambda_{a,1}j}
                  \ket{j,\,
                        j\oplus\lambda_{a,2},\ldots,
                        j\oplus\lambda_{a,n_a}}_{\pi_a},
                  \label{eq:block_GHZ_vector}
                \end{equation}
                while for a singleton block we use the computational basis
                \(\ket{\phi_{\pi_a}^{\lambda}}=\ket{\lambda}_{\pi_a}\).  The
                tensor-product basis
                \begin{equation}
                  \mathcal{G}_\pi
                  \coloneqq
                  \left\{
                  \ket{\Phi_{\boldsymbol{\Lambda}}^\pi}
                  =
                  \bigotimes_{a=1}^{\ell}
                  \ket{\phi_{\pi_a}^{\Lambda_a}}
                  \right\}_{\boldsymbol{\Lambda}}
                \end{equation}
                defines the \(\pi\)-GHZ simplex
                \begin{equation}
                  \Delta_\pi^{\mathrm{GHZ}}
                  \coloneqq
                  \left\{
                  \sum_{\boldsymbol{\Lambda}}
                  p_{\boldsymbol{\Lambda}}^{(\pi)}
                  \ket{\Phi_{\boldsymbol{\Lambda}}^\pi}
                  \!\bra{\Phi_{\boldsymbol{\Lambda}}^\pi}
                  \;\middle|\;
                  p_{\boldsymbol{\Lambda}}^{(\pi)}\geq0,\;
                  \sum_{\boldsymbol{\Lambda}}
                  p_{\boldsymbol{\Lambda}}^{(\pi)}=1
                  \right\}.
                \end{equation}
                Each vertex of \(\Delta_\pi^{\mathrm{GHZ}}\) is product across the
                blocks of \(\pi\); hence
                \(\Delta_\pi^{\mathrm{GHZ}}\subseteq\calD_\pi\).  These partitioned
                simplices interpolate between the computational-basis simplex for the
                finest partition and the global GHZ simplex \(\Delta_{d^N-1}\) for
                the coarsest partition.
        
                They also give simple hierarchy tests.  If a partition
                \(\sigma\) refines \(\pi\), write
                \[
                  \sigma\preceq\pi,
                \]
                and let \(m_\pi(\sigma)\) be the number of blocks of \(\pi\) that are
                split into at least two blocks by \(\sigma\).  Then every
                \(\rho\in\calD_\sigma\) obeys
                \begin{equation}
                  p_{\boldsymbol{\Lambda}}^{(\pi)}(\rho)
                  \coloneqq
                  \bra{\Phi_{\boldsymbol{\Lambda}}^\pi}
                  \rho
                  \ket{\Phi_{\boldsymbol{\Lambda}}^\pi}
                  \leq d^{-m_\pi(\sigma)}
                  \quad
                  \text{for all }\boldsymbol{\Lambda}.
                  \label{eq:pi_probability_bound}
                \end{equation}
                Thus a single \(\pi\)-GHZ outcome above this threshold rules out
                membership in \(\calD_\sigma\).  In the fully separable case
                \(\sigma=\{\{1\},\ldots,\{N\}\}\), the exponent is simply the number
                of nonsingleton blocks of \(\pi\).
        
                A complementary obstruction comes from positivity under partial
                transpose.  If a state is separable across a cut, its partial transpose
                with respect to one side of that cut is positive semidefinite
                \cite{horodecki2009}.  For GHZ-diagonal states this condition reduces
                to small blocks.  For \(N=3\), write
                \[
                  \rho_P=\sum_{\lambda_1,a,b\in\Z_d}
                  p_{\lambda_1,a,b}\,
                  \Phi_{\lambda_1,a,b}
                \]
                and define partial Fourier slices
                \begin{equation}
                  q_t(a,b)
                  \coloneqq
                  \sum_{\lambda_1\in\Z_d}
                  p_{\lambda_1,a,b}\,\om^{t\lambda_1},
                  \qquad t,a,b\in\Z_d.
                  \label{eq:ppt_partial_fourier_slice}
                \end{equation}
                For \(t=0\) this is the fiber mass~\eqref{eq:fiber_mass}.
                For \(u,c\in\Z_d\), let \(\mathsf{P}_{u,c}\) be the \(d\times d\) matrix
                \begin{equation}
                  (\mathsf{P}_{u,c})_{xy}
                  =
                  \frac{1}{d}\,
                  q_{y-x}(u-x-y,\;u-x-y+c),
                  \qquad x,y\in\Z_d,
                  \label{eq:ppt_block_n3}
                \end{equation}
                with all arguments understood modulo \(d\).  Then
                \begin{equation}
                  \rho_P^{T_1}\geq0
                  \quad\Longleftrightarrow\quad
                  \mathsf{P}_{u,c}\succeq0
                  \quad
                  \text{for all }u,c\in\Z_d.
                  \label{eq:ppt_block_criterion_n3}
                \end{equation}
                The criteria for \(\rho_P^{T_2}\) and \(\rho_P^{T_3}\) are obtained by
                permuting the parties.  In particular, every \(2\times2\) principal minor gives the
                scalar necessary condition
                \begin{multline}
          \left| q_{y-x}(u-x-y,\;u-x-y+c) \right|^2 \\
          \leq q_0(u-2x,\;u-2x+c)\, q_0(u-2y,\;u-2y+c).
          \label{eq:ppt_minor_n3}
        \end{multline}
                Violation of any such inequality guarantees non-PPT (and hence non-separability) across the tested
                one-vs-rest bipartition.  Here \(u,c,x,y\in\Z_d\).
        
                The same reduction holds for every $N$: the partial transpose $\rho_P^{T_1}$
                decomposes into $d^{N-1}$ blocks of size $d\times d$, built from the partial
                Fourier transforms
                $q_t(\mathbf a)\coloneqq\sum_{\lambda_1\in\Z_d}p_{\lambda_1,\mathbf a}\om^{t\lambda_1}$
                (Appendix~\ref{app:ppt_general}). Therefore each one-vs-rest PPT test reduces from a
                \(d^N\times d^N\) positivity problem to \(d^{N-1}\) positivity tests of
                \(d\times d\) matrices.  These conditions are necessary for
                separability across the tested cut; their violation gives an analytic
                sufficient entanglement criterion, while positivity alone is not sufficient in
                general.

    \section{Entanglement properties of the GHZ simplex}\label{sec:ent-prop}
    In this section, we construct explicit strictly $k$-separable states for all $k\in\{1,\dots,N\}$ and explicit families of fully separable states inside the GHZ simplex.
                
    \subsection{Depolarization Construction: Fully Separable, Biseparable, and \(k\)-Separable States}
                \label{subsec:depolarization}
                
                We use the channel $\mathcal{E}$ from Sec.~\ref{subsec:channel} to construct explicit examples of each separability class within the GHZ simplex.

                We first introduce the common construction.  For $2\leq m\leq N-1$ set
                \begin{equation}
                  S_m\coloneqq\{m,m+1,\ldots,N\},
                  \quad
                  \pi_m\coloneqq\{\{1\},\ldots,\{m-1\},S_m\}.
                \end{equation}
                Define
                \begin{align}
                  \ket{\eta_m}
                  &\coloneqq
                  \ket{+}_1\otimes\cdots\otimes\ket{+}_{m-1}
                  \otimes\ket{\phi^+}_{S_m}, \notag\\
                  \ket{+}_r
                  &\coloneqq
                  \frac{1}{\sqrt d}\sum_{a\in\Z_d}\ket{a}_r,
                  \qquad
                  \ket{\phi^+}_{S_m}
                  \coloneqq
                  \frac{1}{\sqrt d}\sum_{j\in\Z_d}\ket{j}^{\otimes |S_m|}.
                  \label{eq:eta_m_definition}
                \end{align}
                The vector $\ket{\eta_m}$ is product across the $m$-partition
                $\pi_m$.  For
                $\alpha=(\alpha_2,\ldots,\alpha_{m-1})\in\Z_d^{m-2}$ and
                $t\in\Z_d$ (with empty $\alpha$ when $m=2$), define
                \begin{equation}
                  \Lambda_m(\alpha,t)
                  \coloneqq
                  (0,\alpha_2,\ldots,\alpha_{m-1},t,\ldots,t).
                \end{equation}
                Expanding Eq.~\eqref{eq:GHZ_state} gives
                \begin{equation}
                  \ket{\eta_m}
                  =
                  \frac{1}{\sqrt{d^{m-1}}}
                  \sum_{\alpha\in\Z_d^{m-2}}\sum_{t\in\Z_d}
                  \ket{\phi^{\Lambda_m(\alpha,t)}}.
                \end{equation}
                Therefore the depolarized state is
                \begin{equation}
                  \rho_m
                  \coloneqq
                  \mathcal{E}
                  \left(\ket{\eta_m}\!\bra{\eta_m}\right)
                  =
                  \frac{1}{d^{m-1}}
                  \sum_{\alpha\in\Z_d^{m-2}}\sum_{t\in\Z_d}
                  \Phi_{\Lambda_m(\alpha,t)}.
                  \label{eq:strict_k_GHZ_point}
                \end{equation}
                Since $\mathcal{E}$ preserves separability across any partition
                (Sec.~\ref{subsec:channel}),
                $\rho_m$ is separable across $\pi_m$, hence
                $\rho_m\in\calD_{m\text{-}\sep}$.
                
                \paragraph{A strictly biseparable depolarized point.}
                Let us first see the construction in the biseparable case $m=2$.  Then
                $S_2=\{2,\ldots,N\}$, $\pi_2=\{\{1\},S_2\}$, and
                $\ket{\eta_2}=\ket{+}_1\otimes\ket{\phi^+}_{2\ldots N}$.  With
                $\Lambda_2(t)=(0,t,\ldots,t)$, Eq.~\eqref{eq:strict_k_GHZ_point}
                becomes
                \begin{equation}
                  \rho_{\mathrm{sBS}}
                  \coloneqq
                  \rho_2
                  =
                  \frac{1}{d}\sum_{t\in\Z_d}\Phi_{\Lambda_2(t)}.
                  \label{eq:strict_bs_GHZ_point}
                \end{equation}
                Thus $\rho_{\mathrm{sBS}}\in\BS$.  After partial transposition with
                respect to subsystem $2$, the subspace spanned by
                $\ket{0,1,0,\ldots,0}$ and
                $\ket{1,0,1,\ldots,1}$ contains the block
                \begin{equation}
                  \frac{1}{d^2}
                  \begin{pmatrix}
                    0 & 1\\
                    1 & 0
                  \end{pmatrix},
                  \label{eq:sbs_pt_block}
                \end{equation}
                        which has eigenvalue $-1/d^2$. Since every restriction of a positive
                semidefinite operator is positive semidefinite, $\rho_{\mathrm{sBS}}^{T_2}\not\geq0$,
                and by the PPT criterion $\rho_{\mathrm{sBS}}\notin\FS$.  This gives a strictly
                biseparable GHZ-diagonal point.  For $N=3$, it reduces to
                $\frac12(\Phi_{0,0,0}+\Phi_{0,1,1})$ for $d=2$ and to
                \begin{equation}
                  \rho_{\mathrm{sBS}}^{3,3}
                  =
                  \frac13
                  \left(
                  \Phi_{0,0,0}
                  +
                  \Phi_{0,1,1}
                  +
                  \Phi_{0,2,2}
                  \right)
                \end{equation}
                for $d=3$.
                                The qubit example has $p_{\max}=1/2$ and thus lies on the exact
                biseparability boundary of Ref.~\cite{guhne2010genuine}.
              
                \paragraph{Strictly \(k\)-separable depolarized points.}
                The same construction extends to the full hierarchy: for $2\leq k\leq N-1$,
                $\rho_k\in\calD_{k\text{-}\sep}$ by construction. Moreover, every
                $(k+1)$-partition splits the block $S_k$, and every vector in the support of
                $\rho_k$ has Schmidt rank at least $d$ across any such split
                (Appendix~\ref{app:strict_k}). Hence $\rho_k\notin\calD_{(k+1)\text{-}\sep}$,
                and $\rho_k$ is strictly $k$-separable.

                Applying Sec.~\ref{subsec:channel} to the partition into single
                parties gives $\mathcal{E}(\Pi)\in\FS$ for every pure product projector
                \begin{equation}
                  \Pi=\bigotimes_{s=1}^N\ket{\psi_s}\!\bra{\psi_s}.
                  \label{eq:pure_product_projector}
                \end{equation}
                Let $\mathcal P_{\mathrm{prod}}$ denote the set of all pure product projectors.
                
                \paragraph{Exact fully separable intersection.}
                Define the fully separable part of the GHZ simplex by
                \begin{equation}
                  K_{\FS}^{\mathrm{GHZ}}
                  \coloneqq\FS\cap\Delta_{d^N-1}.
                \end{equation}
                It has the exact convex-hull description
                \begin{equation}
                  K_{\FS}^{\mathrm{GHZ}}
                  =\operatorname{conv}\!\left(
                    \mathcal{E}(\mathcal P_{\mathrm{prod}})
                  \right).
                  \label{eq:exact_GHZ_fs_convex_hull}
                \end{equation}
                One inclusion follows from the preceding separability argument.  Conversely,
                let $\sigma\in K_{\FS}^{\mathrm{GHZ}}$.  Full separability gives a decomposition
                    $\sigma=\sum_r q_r\Pi_r$ into pure product projectors.  Since $\sigma$ is
                already GHZ diagonal, it is a fixed point of the depolarization, and linearity gives
                \begin{equation}
                  \sigma=\mathcal{E}(\sigma)
                  =\sum_r q_r\mathcal{E}(\Pi_r),
                \end{equation}
                which proves the reverse inclusion.  By Carath\'eodory's theorem, no more than
                $d^N$ product-depolarized generators are required, because the GHZ simplex
                has affine dimension $d^N-1$.

                To make the generators in Eq.~\eqref{eq:exact_GHZ_fs_convex_hull} explicit,
                write a pure state as $\ket{\Psi}=\sum_{i_1,\ldots,i_N}\Psi_{i_1\ldots i_N}\ket{i_1,\ldots,i_N}$.
                By Eq.~\eqref{eq:GHZ_state},
                $\braket{\phi^\Lambda}{\Psi}=d^{-1/2}\sum_j\om^{-\lambda_1 j}\Psi_{j,\,j\oplus a_2,\ldots,\,j\oplus a_N}$,
                so $\mathcal E(\ket{\Psi}\!\bra{\Psi})=\sum_\Lambda p_\Lambda\Phi_\Lambda$ with
                \begin{equation}
                  p_{\lambda_1,\mathbf a}
                  =\frac1d\left|\sum_{j=0}^{d-1}\om^{-\lambda_1 j}\,
                     \Psi_{j,\,j\oplus a_2,\ldots,\,j\oplus a_N}\right|^2 .
                  \label{eq:product_depolarization_probability}
                \end{equation}
                This holds for every pure state. For a product input~\eqref{eq:pure_product_projector}
                the coefficients factorize, $\Psi_{i_1\ldots i_N}=\prod_{s=1}^N\braket{i_s}{\psi_s}$.
                Consequently, a GHZ probability vector is fully separable if and only if it is a
                convex combination of probability vectors~\eqref{eq:product_depolarization_probability}
                with factorizing coefficients. Restricting the generating family to finitely many product inputs gives polytopal inner approximations. The convex-hull formula itself does not determine whether the full region is a polytope.

                As a simple subfamily that is fully separable, take a projector onto the computational basis state
                $\ket{x_1,\ldots,x_N}\!\bra{x_1,\ldots,x_N}$ and set
                $a_s=x_s-x_1\pmod d$ for $s\geq2$.  Its depolarization is the fiber state
                \begin{align}
                  \eta_{\mathbf a}
                  &\coloneqq\frac{1}{d}\sum_{\lambda_1\in\Z_d}
                  \Phi_{\lambda_1,\mathbf a} \notag\\
                  &=\frac{1}{d}\sum_{j=0}^{d-1}
                  \ket{j,j\oplus a_2,\ldots,j\oplus a_N}
                  \!\bra{j,j\oplus a_2,\ldots,j\oplus a_N},
                  \label{eq:fiber_state}
                \end{align}
                i.e., the uniform distribution on the fiber of $\mathbf a$.
                The second line follows from
                $\sum_{\lambda_1\in\Z_d}\om^{\lambda_1(j-k)}=d\,\delta_{jk}$; it is an
                explicit mixture of product projectors, so each
                $\eta_{\mathbf a}$ is fully separable.  The $d^{N-1}$ fiber states have disjoint supports
                in the GHZ probability coordinates; therefore, their convex hull, the
                \emph{fiber simplex} $\mathcal P_{\mathrm{fib}}$, is a
                $(d^{N-1}-1)$-dimensional FS simplex inside
                $K_{\FS}^{\mathrm{GHZ}}$.
                \paragraph{Fully separable states in the GHZ simplex.}
            The fiber states are the simplest such subfamily. For the
            figure and for numerical experiments, we use further explicit
            finite subfamilies of product-depolarized generators.  A second one is obtained by
            applying local Fourier transforms to computational-basis product states.
            Let
            \begin{equation}
              F\ket{r}=\frac{1}{\sqrt d}\sum_{x\in\Z_d}\om^{rx}\ket{x}.
              \label{eq:local_fourier}
            \end{equation}
            For the product input $F\ket{r_1}\otimes\cdots\otimes F\ket{r_N}$, the sum in
            Eq.~\eqref{eq:product_depolarization_probability} is a geometric sum over $j$, and
            \begin{equation}
              p_{\lambda_1,\mathbf a}
              =
              \begin{cases}
                d^{1-N},&
                \lambda_1\equiv r_1+\cdots+r_N \pmod d,\\
                0,&
                \text{otherwise}.
              \end{cases}
              \label{eq:fourier_orbit_full_S}
            \end{equation}
            i.e., a uniform distribution over the $d^{N-1}$ GHZ basis states with fixed
            phase label. This is one case of a larger family of partial-Fourier inputs,
            treated in Appendix~\ref{app:partial_fourier}.

\paragraph{Fiber and Fourier directions.}
The all-Fourier
inputs of Eq.~\eqref{eq:fourier_orbit_full_S} give the phase layers
\begin{equation}
 L_c\coloneqq d^{1-N}\sum_{\mathbf a\in\Z_d^{N-1}}\Phi_{c,\mathbf a},
 \qquad c=r_1+\cdots+r_N\in\Z_d.
 \label{eq:general_phase_layer}
\end{equation}
i.e., the uniform distributions on all GHZ basis states with phase label $c$.
Each layer has fiber mass $q_0(\mathbf a)=d^{1-N}$ in every fiber, as does
$\rho_{\mathrm{mix}}=\id/d^N$, which is the barycenter of both the layers and the
fiber simplex. The two families therefore move the state in complementary
directions: the fiber states change the fiber masses ($d^{N-1}-1$ independent
directions), while the differences $L_c-\rho_{\mathrm{mix}}$ keep all fiber masses
fixed and only redistribute probability along the phase label ($d-1$ independent
directions, since $\sum_c(L_c-\rho_{\mathrm{mix}})=0$). Hence
\begin{equation}
 \dim\operatorname{aff}\bigl(\{\eta_{\mathbf a}\}_{\mathbf a}\cup\{L_c\}_{c\in\Z_d}\bigr)
 =d^{N-1}+d-2.
 \label{eq:general_fiber_fourier_dimension}
\end{equation}
Inputs with at least one computational-basis factor give nothing new: that factor
fixes the summation index $j$ in Eq.~\eqref{eq:product_depolarization_probability},
so the distribution is uniform in the phase label within each fiber and lies in
the fiber simplex (Appendix~\ref{app:partial_fourier}).

\subsection{Finite phases and full-dimensional fully separable simplices}
\label{sec:finite_phase_span}
The fiber states change the fiber masses. To reach the remaining directions,
which keep all fiber masses fixed, we use equal-modulus product inputs with
discrete local phases. For an integer $m$ and $\zeta=e^{2\pi i/m}$, let
\begin{equation}
 |\psi_s(\mathbf k_s)\rangle=\frac1{\sqrt d}\sum_{i=0}^{d-1}\zeta^{k_{s,i}}|i\rangle,
 \label{eq:general_finite_phase_inputs}
\end{equation}
where $s=1,\ldots,N$ labels the subsystem,
$\mathbf k_s=(0,k_{s,1},\ldots,k_{s,d-1})\in\{0\}\times\Z_m^{d-1}$ lists its local
phases, and $k_{s,0}=0$ removes an irrelevant global phase. Let
$|\Psi_{\mathbf k}\rangle=\bigotimes_{s}|\psi_s(\mathbf k_s)\rangle$.
Every coefficient of $|\Psi_{\mathbf k}\rangle$ has modulus $d^{-N/2}$, so
Eq.~\eqref{eq:product_depolarization_probability} and Parseval's identity give
\begin{equation}
 q_0(\mathbf a)=d^{1-N}\qquad\text{for every }\mathbf a.
 \label{eq:general_uniform_fiber_mass}
\end{equation}
All these states therefore lie in the \emph{uniform-fiber slice}, the set of
GHZ-diagonal states whose fiber masses all equal $d^{1-N}$. The slice is fixed by
$d^{N-1}$ independent linear constraints and has dimension
$d^N-d^{N-1}=(d-1)d^{N-1}$, independently of $m$. The question is whether the
phase family fills it.

\begin{theorem}[Finite phase span and full-dimensional FS simplex]
\label{thm:general_finite_phase_simplex}
Let $N\ge2$, $d\ge2$ and $m\ge3$. The depolarizations of the $m^{N(d-1)}$ product
inputs $|\Psi_{\mathbf k}\rangle$ have affine hull equal to the uniform-fiber slice,
of dimension $(d-1)d^{N-1}$. Moreover, the $d^{N-1}$ fiber states together with
suitably chosen $(d-1)d^{N-1}$ of these states are the vertices of a fully
separable simplex of dimension $d^N-1$.
\end{theorem}

The proof is in Appendix~\ref{app:general_finite_phase}. Each term of Eq.~\eqref{eq:product_depolarization_probability}, after expanding the
square, is a product of single-site factors $\zeta^{k_{s,i}-k_{s,l}}$, and for
$m\ge3$ these distinguish the ordered pair $(i,l)$ from $(l,i)$ because
$\zeta\neq\bar\zeta$. This is why the condition is $m\ge3$ and why it does not
depend on $d$. Any $m\ge3$ works; a natural choice is $m=d$ for $d\ge3$ (for $d=3$
this is the family of Sec.~\ref{subsec:example_qutrit}) and $m=3$ for qubits.

For $m=2$ the condition fails. All amplitudes are then real, and
Eq.~\eqref{eq:product_depolarization_probability} gives
$p_{\lambda_1,\mathbf a}=p_{-\lambda_1,\mathbf a}$, which confines the family to a
proper subspace of the slice. For qubits the inputs are just $F|0\rangle$ and
$F|1\rangle$, and the family consists only of $L_0$ and $L_1$.

The simplex part of the theorem is an existence statement;
Sec.~\ref{subsec:example_qutrit} gives an explicit selection for three qutrits.

            \subsection{Examples: Fully Separable States for $N=3$, $d=2$}
            \label{subsec:example_qubit}
                        Full separability in this simplex is already characterized by
            the results of G\"uhne and Han and Kye
            \cite{guhne2011full,han2017separability}. The families below give
            explicit separable generators and geometric inner
            approximations within that characterized region.
            \paragraph{Depolarized pure product states.}
            Every one-qubit pure state can, up to a phase, be written as
            \begin{equation}
              \ket{\psi_s}
              =
              \cos\frac{\beta_s}{2}\ket0
              +
              e^{i\theta_s}\sin\frac{\beta_s}{2}\ket1,
              \label{eq:bloch_qubit}
            \end{equation}
            with $\beta_s\in[0,\pi]$, $\theta_s\in[0,2\pi)$, where $s=1,2,3$ labels
            the qubit. We write the GHZ projectors as
            $\Phi_{\lambda,a,b}$, $\lambda,a,b\in\Z_2$. For $d=2$ the sum in
            Eq.~\eqref{eq:product_depolarization_probability} has only the terms $j=0,1$,
            and with $\bar a\coloneqq 1-a$, $\bar b\coloneqq 1-b$ it gives
            \begin{equation}
              p_{\lambda,a,b}=\frac12\left|\Psi_{0ab}+(-1)^\lambda\,\Psi_{1\bar a\bar b}\right|^2,
              \label{eq:bloch_general_probability}
            \end{equation}
            where $\Psi_{x_1x_2x_3}=\prod_{s=1}^3\braket{x_s}{\psi_s}$, with
            $\braket{0}{\psi_s}=\cos(\beta_s/2)$ and $\braket{1}{\psi_s}=e^{i\theta_s}\sin(\beta_s/2)$.
            Equation~\eqref{eq:bloch_general_probability} is the most general GHZ
            probability vector obtainable by depolarizing a three-qubit product state,
            with $(\beta_1,\theta_1,\beta_2,\theta_2,\beta_3,\theta_3)\in([0,\pi]\times[0,2\pi))^3$;
            each such point is FS.
    
            Taking $\beta_s\in\{0,\pi\}$ for every $s$ gives computational-basis inputs and
            hence the fiber states $\eta_{\mathbf a}$.
     
            \paragraph{The equatorial family.}
            Setting $\beta_s=\pi/2$ for every $s$ gives the equatorial product inputs (the qubit
            $T$-after-Hadamard family),
            \begin{equation}
              \ket{\theta_s}
              =
              \frac{\ket0+e^{i\theta_s}\ket1}{\sqrt2}.
              \label{eq:equatorial_product}
            \end{equation}
            Their GHZ probabilities follow directly from
            Eq.~\eqref{eq:bloch_general_probability}:
            \begin{equation}
              p_{\lambda,a,b}
              =
              \frac18
              \left[
              1+
              (-1)^\lambda
              \cos\left(
              \theta_1+(1-2a)\theta_2+(1-2b)\theta_3
              \right)
              \right].
              \label{eq:equatorial_probability}
            \end{equation}
            
            \paragraph{Inner approximation of the FS set.}
            The four fiber states of Eq.~\eqref{eq:fiber_state} are
            $F_{a,b}\coloneqq\tfrac12(\Phi_{0,a,b}+\Phi_{1,a,b})$, $a,b\in\Z_2$. 
            
            
            The layers $L_0,L_1$ of Eq.~\eqref{eq:general_phase_layer}, i.e., uniform
            weight $1/4$ on the four GHZ vertices with $\lambda=0$ or $\lambda=1$, are the
            depolarizations of $\ket{+++}$ and $\ket{---}$, i.e., the equatorial points
            $(0,0,0)$ and $(\pi,\pi,\pi)$.
            
\begin{figure}[h]
            \centering
            \includegraphics[width=0.48\textwidth]{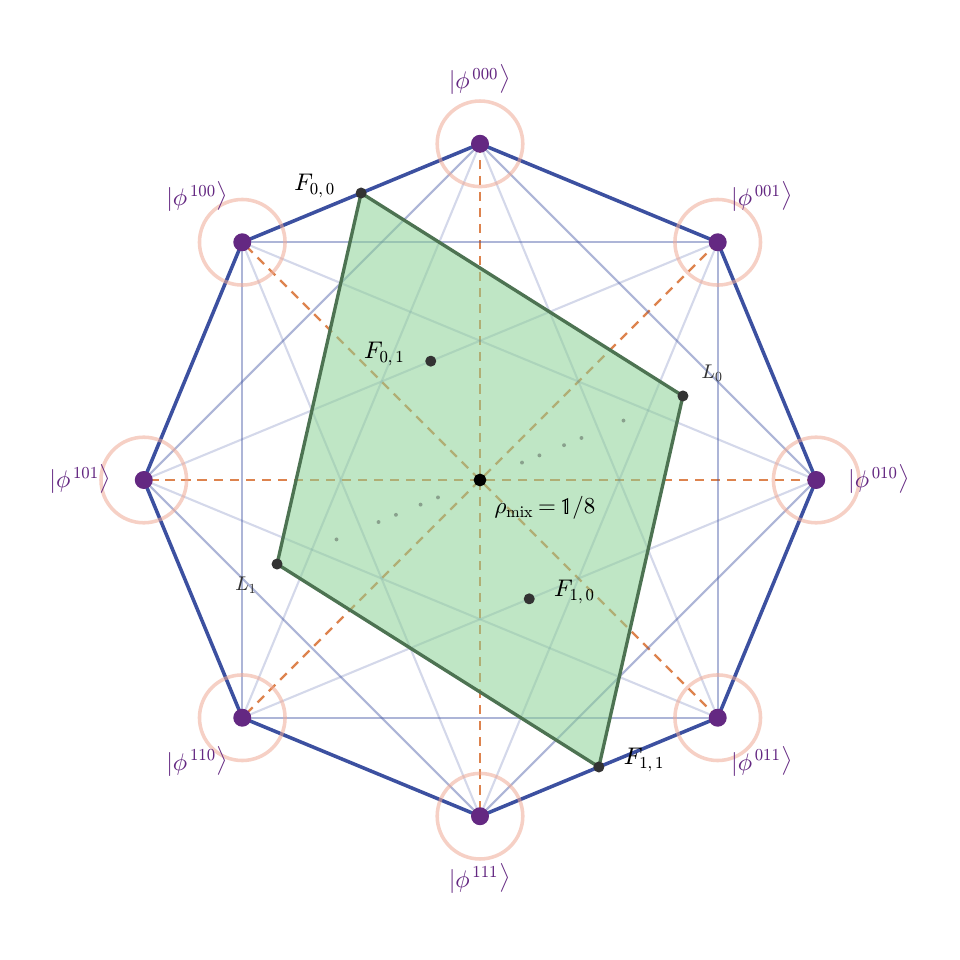}
            \caption{\emph{Planar shadow of the seven-dimensional
            GHZ simplex $\Delta_7$ for $N=3$ qubits.} The eight GHZ vertices $\ket{\phi^\Lambda}$ sit at the vertices
            $v_\Lambda$ of a regular octagon in bit-complement order, joined by
            all $\binom82=28$ simplex edges. Dashed diagonals connect
            bit-complement pairs through the centre, where the maximally
            mixed reference state $\rho_{\mathrm{mix}}=\mathbbm{1}/8$ sits.
            Each probability vector is projected linearly as
            $\mathbf p\mapsto\sum_\Lambda p_\Lambda v_\Lambda$.
            The green polygon is the convex hull of the projected fiber
            states $F_{0,0},F_{0,1},F_{1,0},F_{1,1}$ together with the equatorial
            family of Eq.~\eqref{eq:equatorial_probability}, sampled at
            $\theta_s\in\{0,\pi/4,\ldots,7\pi/4\}$ (faint dots). The labelled points $L_0$ and $L_1$ are the special
            equatorial cases $(0,0,0)$ and $(\pi,\pi,\pi)$, respectively;
            they are already part of the sampled family. All states used
            to construct the hull are fully separable, being depolarizations of
            product inputs. The polygon is an inner approximation of the projected FS set. The circles around the GHZ vertices
            schematically illustrate the GME criterion
            $\max_\Lambda p_\Lambda>1/2$
            (Sec.~\ref{subsec:majorization}); they are schematic.}
            \label{fig:shadow_delta7}
            \end{figure}

            After applying the planar projection used in Fig.~\ref{fig:shadow_delta7}, the
            convex hull of their images gives the green inner shadow. Every point of the green polygon has at least one fully separable preimage.

\subsection{Three qutrits: a full-dimensional fully separable simplex}
\label{subsec:example_qutrit}

For $N=d=3$, write the GHZ projectors as $\Phi_{\lambda_1,a,b}$,
with $\lambda_1,a,b\in\Z_3$. The constructions below progress from
varying fiber masses to redistributing phase probabilities within each
fiber, and finally to an explicit simplex with $27$ FS vertices.
The Fourier matrix of Eq.~\eqref{eq:local_fourier} becomes
\begin{equation}
 F=\frac1{\sqrt3}
 \begin{pmatrix}1&1&1\\1&\om&\om^2\\1&\om^2&\om\end{pmatrix},
 \qquad \om=e^{2\pi i/3}.
 \label{eq:Fourier_basis}
\end{equation}
For a normalized product input $\bigotimes_{s=1}^3|\psi_s\rangle$,
Eq.~\eqref{eq:product_depolarization_probability} reads
\begin{equation}
 p_{\lambda_1,a,b}=\frac13\left|
 \sum_{j=0}^2\om^{-\lambda_1j}
 \langle j|\psi_1\rangle
 \langle j\oplus a|\psi_2\rangle
 \langle j\oplus b|\psi_3\rangle
 \right|^2.
 \label{eq:master_product_formula}
\end{equation}

\begin{figure*}[t]
\centering
\begin{minipage}[t]{0.31\textwidth}
  \centering
  \textbf{(a)} $\eta_{0,0}$\par\medskip
  \fibersupportpanel{0}\par\smallskip
  \small One occupied fiber\\
  $p=1/3$ at each active point
\end{minipage}\hfill
\begin{minipage}[t]{0.31\textwidth}
  \centering
  \textbf{(b)} $L_0$\par\medskip
  \fibersupportpanel{1}\par\smallskip
  \small One fixed phase layer\\
  $p=1/9$ at each active point
\end{minipage}\hfill
\begin{minipage}[t]{0.31\textwidth}
  \centering
  \textbf{(c)} $\rho^{\mathrm{ph}}_{0,1,1}$\par\medskip
  \fibersupportpanel{2}\par\smallskip
  \small $\lambda_1=a+b\pmod3$\\
  $p=1/9$ at each active point
\end{minipage}

\medskip
\begin{tikzpicture}[font=\small]
  \foreach \phase/\phasecolor/\pos in
    {0/phasezero/0,1/phaseone/2.3,2/phasetwo/4.6}{%
    \fill[\phasecolor] (\pos,0) circle[radius=2.6pt];
    \node[anchor=west] at (\pos+0.16,0) {$\lambda_1=\phase$};
  }
  \fill[black!20] (6.9,0) circle[radius=1.6pt];
  \node[anchor=west] at (7.06,0) {$p=0$};
\end{tikzpicture}

\caption{Support of fully separable three-qutrit
families.  Each grid cell is one fixed $(a,b)$ fiber, with phase positions
$\lambda_1=0$ (top), $1$ (lower left), and $2$ (lower right).
Gray points have zero probability.  Active marker areas are proportional
to the probabilities: panel (a) has three entries of weight $1/3$,
while panels (b) and (c) each have nine entries of weight $1/9$.
In panel (c), the occupied phases by row are $(0,1,2)$, $(1,2,0)$,
and $(2,0,1)$.}
\label{fig:fiber_support_dimension}
\end{figure*}

\subsubsection{Construction I: nine rank-3 fiber states}

For $N=d=3$, Eq.~\eqref{eq:fiber_state} gives the nine rank-$3$ fiber states
\begin{align}
 \eta_{a,b}&\coloneqq\frac13\sum_{\lambda_1=0}^2\Phi_{\lambda_1,a,b}
 \label{eq:eta_fibers}\\
 &=\frac13\sum_{j=0}^2
 |j,j\oplus a,j\oplus b\rangle\langle j,j\oplus a,j\oplus b|,
 \qquad a,b\in\Z_3.
 \label{eq:eta_product_decomp}
\end{align}
The fiber simplex $\mathcal P_{\mathrm{fib}}$ is then an $8$-simplex with barycenter
\begin{equation}
 \frac19\sum_{a,b}\eta_{a,b}
 =\frac1{27}\sum_{\lambda_1,a,b}\Phi_{\lambda_1,a,b}
 =\rho_{\mathrm{mix}}=\frac{\id}{27}.
 \label{eq:maxmix_from_fibers}
\end{equation}

\subsubsection{Construction II: three Fourier phase layers}

For $N=d=3$, the $27$ Fourier inputs $F|r_1\rangle\otimes F|r_2\rangle\otimes F|r_3\rangle$
yield, by Eq.~\eqref{eq:fourier_orbit_full_S}, only the three layers
$L_c=\frac19\sum_{a,b\in\Z_3}\Phi_{c,a,b}$ of Eq.~\eqref{eq:general_phase_layer},
with $c=r_1+r_2+r_3 \bmod 3$. By Eq.~\eqref{eq:general_fiber_fourier_dimension},
the fiber states and these layers span an affine space of dimension $8+2=10$,
below the ambient dimension $26$.

\subsubsection{Construction III: finite local phases and affine-graph states}
The nine local phase states can be written using the Fourier matrix and
a diagonal phase gate: Write $D\coloneqq\operatorname{diag}(1,\om,\om)$, and for $r,s\in\mathbb Z_3$, define
\begin{equation}
\begin{aligned}
 |u_{r,s}\rangle\coloneqq& D^{2r-s}F|s-r\rangle\\
 =&\frac1{\sqrt3}\bigl(|0\rangle+\om^r|1\rangle+\om^s|2\rangle\bigr).
\end{aligned}
\label{eq:u_rs}
\end{equation}
Thus these states are exactly $D^\mu F|\nu\rangle$, with
$\mu,\nu\in\Z_3$.  They are not all powers $F^\mu|\nu\rangle$:
$F^2|\nu\rangle=|-\nu\rangle$ (mod $3$) and $F^4=\id$, so powers of $F$
give only the computational and ordinary Fourier bases.
The labels $(r,s)$ are retained below to specify the product inputs explicitly.

Define the product state
\begin{equation}
|\Psi^{\mathrm{ph}}_{\mathbf r,\mathbf s}\rangle
=|u_{r_1,s_1}\rangle\otimes
 |u_{r_2,s_2}\rangle\otimes
 |u_{r_3,s_3}\rangle
\end{equation}
and its GHZ depolarization
\begin{equation}
    \rho^{\mathrm{ph}}_{\mathbf r,\mathbf s}
    \coloneqq\mathcal E\bigl(\ghzprojector{\Psi^{\mathrm{ph}}_{\mathbf r,\mathbf s}}\bigr).
\end{equation}
Depolarizing the $9^3=729$ product inputs gives only $28$ distinct fully
separable states. Twenty-seven of them are the \emph{affine-graph states}
\begin{equation}
\rho^{\mathrm{ph}}_{c,u,v}
\coloneqq\frac19\sum_{a,b\in\Z_3}\Phi_{c+ua+vb,a,b},
\qquad c,u,v\in\Z_3,
\label{eq:affine_graph_state}
\end{equation}
each obtained from nine inputs; the remaining $486$ inputs all give
$\rho_{\mathrm{mix}}$. Here, $c$ is the phase offset and $(u,v)$ are the slopes
across the fiber coordinates; the Fourier layers are the case
$\rho^{\mathrm{ph}}_{c,0,0}=L_c$. One product input for each graph state is
\begin{equation}
\begin{gathered}
|\Psi^{\mathrm{ph}}_{c,u,v}\rangle
=D^{u+v}F|c\rangle\otimes D^{-u}F|0\rangle
\otimes D^{-v}F|0\rangle,\\
\rho^{\mathrm{ph}}_{c,u,v}
=\mathcal E\bigl(\ghzprojector{\Psi^{\mathrm{ph}}_{c,u,v}}\bigr).
\end{gathered}
\label{eq:graph_product_input}
\end{equation}
The proof in Appendix~\ref{app:three_qutrit_proof} writes each local phase as a
polynomial of degree two over $\Z_3$. The exponent in
Eq.~\eqref{eq:master_product_formula} is then quadratic in $j$. If its quadratic
coefficient vanishes, the sum over $j$ collapses to a Kronecker delta and gives a
graph state; otherwise it is a quadratic Gauss sum of modulus $\sqrt3$ for every
$\lambda_1$, which gives $\rho_{\mathrm{mix}}$.

\paragraph{Continuous phases and the scope of the finite construction.}

For arbitrary equal-modulus local inputs
$3^{-1/2}\sum_{j=0}^2e^{i\theta_{s,j}}|j\rangle$, Parseval's identity
in Eq.~\eqref{eq:master_product_formula} gives for the fiber masses $q_0(a,b)$ of
Eq.~\eqref{eq:fiber_mass}
\begin{equation}
q_0(a,b)=\sum_{\lambda_1=0}^2p_{\lambda_1,a,b}=\frac19.
\label{eq:uniform_fiber_mass}
\end{equation}
Thus every one of the nine fibers carries strictly positive mass.
Their affine hull lies in the $18$-dimensional uniform-fiber slice.
The finite family of Sec.~\ref{sec:finite_phase_span}
already spans this slice; the graph states span it together with
$\rho_{\mathrm{mix}}$, so continuous phases are unnecessary for the
full-dimensional simplex construction.

\paragraph{Minimum rank within the equal-modulus family.}
Equation~\eqref{eq:uniform_fiber_mass} implies that every depolarized
equal-modulus phase state has at least one occupied GHZ projector in each
of the nine fibers. Since these projectors are mutually orthogonal,
$\operatorname{rank}\rho\ge9$.
The affine-graph states have exactly one occupied GHZ vertex per fiber,
each with weight $1/9$, and therefore attain the bound with
$\operatorname{rank}\rho^{\mathrm{ph}}_{c,u,v}=9$.
Minimum rank distinguishes these states within the entire equal-modulus
family, including continuous local phases. Their distribution is as
concentrated as the fixed fiber masses permit. In particular,
$p_{\lambda_1,a,b}\le1/9$ implies
$\operatorname{tr}\rho^2\le1/9$, with equality for the graph states.
They are thus least mixed in the sense of maximal purity within this
family. Their eighteen zero GHZ probabilities place them on the boundary
of $\Delta_{26}$ and of $K_{\FS}^{\mathrm{GHZ}}$, providing sparse vertices
for the FS construction.

\paragraph{Two offsets for each slope.}
For any fixed slope pair $(u,v)$, the three offsets partition the GHZ
vertices, so their average is
$\frac13\sum_{c=0}^2\rho^{\mathrm{ph}}_{c,u,v}
=\frac1{27}\sum_{\lambda_1,a,b}\Phi_{\lambda_1,a,b}
=\rho_{\mathrm{mix}}$.
The centered direction with $c=2$ is therefore the negative sum of those
with $c=0,1$. Only two independent offset directions per slope are needed,
and their common reference $\rho_{\mathrm{mix}}$ already belongs to the
fiber simplex. Taking the same two offsets $c\in\{0,1\}$ for all nine
slope pairs gives $9\times2=18$ phase directions. Any two of the three
offsets could be used consistently; the selection treats all slopes
uniformly and requires no enumeration by rank.

Combining these eighteen minimum-rank graph states with the nine fiber
states defines
\begin{equation}
\begin{gathered}
 \mathcal S_{26}\coloneqq\operatorname{conv}\mathcal V_{26},\\
 \mathcal V_{26}\coloneqq\{\eta_{a,b}:a,b\in\Z_3\}\\
 \quad\cup\{\rho^{\mathrm{ph}}_{c,u,v}:c=0,1,\ u,v\in\Z_3\}.
\end{gathered}
\label{eq:S26}
\end{equation}

\begin{proposition}[Affine independence of the selected vertices]
\label{prop:three_qutrit_simplex}
The 27 states in Eq.~\eqref{eq:S26} are affinely independent.  Consequently $\mathcal S_{26}$ is a $26$-dimensional simplex and
\begin{equation}
    \mathcal S_{26}\subseteq K_{\FS}^{\mathrm{GHZ}}\subseteq\Delta_{26}.
\end{equation}
\end{proposition}

The fiber states control the eight independent fiber masses, while the
graph states supply the remaining eighteen directions at fixed fiber
mass. Appendix~\ref{app:three_qutrit_proof} proves their independence, and
Fig.~\ref{fig:fiber_support_dimension} shows the support patterns of the
three families in GHZ coordinates.

                \section{CONCLUSION AND OUTLOOK}
                \label{sec:Conclusion}
                In this paper, we introduced a systematic and symmetric definition of multipartite GHZ-diagonal states.
                We showed how a decomposition of these states in the Heisenberg--Weyl basis is naturally Fourier-dual to the more conventional probabilistic description in terms of the GHZ basis. Using this picture, we highlighted the symmetry properties of the family of GHZ diagonal states under several finite groups. We constructed a non-entangling depolarizing channel that projects arbitrary density operators onto the set of GHZ diagonal states, providing a bridge between our treatment and more general quantum systems.
                
                Using the group symmetries of the family of GHZ-diagonal states, we derived multipartite separability criteria.
                This led us to the notion of partitioned GHZ diagonal states which are convex combinations of partitioned tensor products of GHZ diagonal states.
                
                With these tools at hand, we showed that the GHZ simplex contains strictly $k$-separable states for any $k\in\{1,\dots,N\}$ by explicitly constructing such families.
    
                                A finite set of equal-modulus local phases already spans
                all directions at fixed fiber mass. Combined with the
                fiber states, this gives a full-dimensional FS simplex
                for every $N$ and $d$ in
                Theorem~\ref{thm:general_finite_phase_simplex}.
                For three qutrits, Proposition~\ref{prop:three_qutrit_simplex}
                makes the selection explicit: nine rank-$3$ fiber states
                and eighteen rank-$9$ affine-graph states give $27$
                affinely independent vertices in $\Delta_{26}$.

    The theoretical framework developed here also provides a basis for training
    machine-learning classifiers, such as support vector machines (SVMs) with
    linear or radial basis function kernels, to investigate the $k$-separability
    hierarchy of GHZ-diagonal states. The GHZ probabilities $p_\Lambda$, the real
    and imaginary parts of their Fourier coefficients $f_M$, the flattened
    correlation-tensor norms of Sec.~\ref{subsec:ccnr}, and the PPT block
    eigenvalues of Sec.~\ref{subsec:partition} can serve as physically motivated
    input features.

    Our analytic results also provide certified labels. The explicit strictly
    $k$-separable constructions give states whose separability class is known in
    advance and can be used to generate training data, and whenever one of the
    sufficient criteria or witnesses applies, it settles the class of the state.
        For three-qubit GHZ-diagonal benchmarks, the exact full-separability
    criterion \cite{guhne2011full,han2017separability} can be combined with
    the exact biseparability condition $p_{\max}\leq1/2$
    \cite{guhne2010genuine} to assign FS, strictly BS, and GME labels
    throughout $\Delta_7$. This provides a fully characterized benchmark
    for classification methods before their application to $d>2$.
    Preliminary numerical studies with three-qutrit GHZ-diagonal states, using
    SVMs with these analytic features, have shown very promising performance in
    distinguishing fully separable, strictly biseparable, and genuinely
    multipartite entangled states \cite{burduli_ml_ghz_in_preparation}.

    Related approaches have already been demonstrated through the numerical construction of entanglement witnesses by Greenwood \emph{et al.}~\cite{greenwood2023witnesses} and the cascaded SVM classification of three-qubit entanglement by Lajevardi \emph{et al.}~\cite{lajevardi2026cascaded}.  Applying such methods to the structures derived here offers a natural numerical extension of our analytical results, enabling the exploration of regions of the GHZ simplex where the present criteria remain inconclusive.

    Several aspects remain to be worked out: More explicit results for the $N$-qubit case would be desirable in light of their application in quantum communication and key distribution protocols. Apart from that, a framework needs to be developed for the case where the local subsystems have unequal dimensions. Furthermore, it would be desirable to treat a countably infinite dimensional analogue, which could be used to describe photons in a cavity.

\appendix
\section{Partial-transpose blocks for general \texorpdfstring{$N$}{N}}
\label{app:ppt_general}
Let $N\geq3$ and let $q_t(\mathbf a)$ be the partial Fourier transforms of
Sec.~\ref{subsec:partition}.
                For
                \(\tau\in\Z_d\) and
                \(\boldsymbol{\delta}=(\delta_2,\ldots,\delta_{N-1})
                \in\Z_d^{N-2}\), set
                \begin{equation}
    \begin{split}
      (\mathsf{P}_{\tau,\boldsymbol{\delta}})_{rs}
      &= \frac{1}{d}\,
      q_{r-s}\bigl(
        \delta_2+r+s-\tau,\ldots, \\
        &\quad \delta_{N-1}+r+s-\tau,
        r+s-\tau
      \bigr).
      \label{eq:ppt_block_general}
    \end{split}
    \end{equation}
                Then
                \begin{equation}
                  \rho_P^{T_1}\geq0
                  \quad\Longleftrightarrow\quad
                  \mathsf{P}_{\tau,\boldsymbol{\delta}}\succeq0
                  \quad
                  \text{for all }\tau\in\Z_d,\;
                  \boldsymbol{\delta}\in\Z_d^{N-2}.
                  \label{eq:ppt_block_criterion_general}
                \end{equation}

\section{Strict \texorpdfstring{$k$}{k}-separability of the depolarized points}
\label{app:strict_k}
We show that $\rho_k$ of Eq.~\eqref{eq:strict_k_GHZ_point} with $m=k$ is not
$(k+1)$-separable.
          Every $(k+1)$-partition
        must split the block $S_k$. Now consider any bipartition $A|A^c$ that
        splits $S_k$. A nonzero vector in the span of the GHZ vectors appearing in
        Eq.~\eqref{eq:strict_k_GHZ_point} with $m=k$ can be grouped according to the
        common computational-basis value on $S_k$ as
        \begin{align}
        |\psi\rangle=\sum_{y\in\mathbb Z_d}|\psi_y\rangle,
        \qquad
        |\psi_y\rangle\in H_A^{(y)}\otimes H_{A^c}^{(y)}.
        \end{align}
        Here, $y$ labels the common value shared by the parties in $S_k$. Since $S_k$
        has at least one party on each side of the split, different values of $y$ give
        orthogonal local supports on both sides:
        \begin{align}
        H_A^{(y)}\perp H_A^{(y')}
        \quad\text{and}\quad
        H_{A^c}^{(y)}\perp H_{A^c}^{(y')}
        \qquad (y\neq y').
        \end{align}
        
        Hence, after tracing out $A^c$, the reduced density operator on $A$ has at
        least $d$ nonzero mutually orthogonal contributions. Its rank, and therefore
        the Schmidt rank, is at least $d$. Thus no such vector is product across the
        split.
        
        If $\rho_k$ admitted a $(k+1)$-separable decomposition, its pure components
        would have to lie in this same span, contradicting the Schmidt-rank argument.
        Thus, $\rho_k\notin\calD_{(k+1)\text{-}\sep}$, so $\rho_k$ is strictly
        $k$-separable.

\section{Partial-Fourier product inputs}
\label{app:partial_fourier}
Here we determine the depolarization of every product input that has a local
Fourier factor $F\ket{r_s}$, Eq.~\eqref{eq:local_fourier}, on a subset $S$ of the
sites and a computational-basis factor $\ket{x_s}$ on the others. Each such
depolarization is uniform on its support.
            For a subset $S\subseteq\{1,\ldots,N\}$, define the product state
            \begin{equation}
              \ket{\Psi_{S,\mathbf x,\mathbf r}}
              =
              \bigotimes_{s\notin S}\ket{x_s}
              \otimes
              \bigotimes_{s\in S}F\ket{r_s},
              \label{eq:fourier_orbit_product}
            \end{equation}
            where $x_s,r_s\in\Z_d$.  Its depolarization
            \begin{equation}
              \mathcal E
              \left(
              \ket{\Psi_{S,\mathbf x,\mathbf r}}
              \bra{\Psi_{S,\mathbf x,\mathbf r}}
              \right)
              =
              \sum_{\Lambda}p^{S,\mathbf x,\mathbf r}_{\Lambda}\Phi_\Lambda
              \label{eq:fourier_orbit_depolarization}
            \end{equation}
            is fully separable.
    
            Using Eq.~\eqref{eq:product_depolarization_probability}, these probabilities are
            explicit:
            \begin{equation}
              p^{S,\mathbf x,\mathbf r}_{\lambda_1,\mathbf a}
              =
              \frac1d
              \left|
              \sum_{j\in\Z_d}
              \om^{-\lambda_1 j}
              \prod_{s\notin S}
              \delta_{j\oplus a_s,x_s}
              \prod_{s\in S}
              \frac{1}{\sqrt d}
              \om^{r_s(j\oplus a_s)}
              \right|^2.
              \label{eq:fourier_orbit_probability}
            \end{equation}
    
            Several simple cases follow immediately.  If $S=\varnothing$, this reduces to
            the fiber states $\eta_{\mathbf a}$.  If $S=\{1,\ldots,N\}$, all local factors are
            Fourier-basis states and the sum over $j$ gives
            Eq.~\eqref{eq:fourier_orbit_full_S}.
            Thus one obtains uniform distributions over $d^{N-1}$ GHZ vertices with fixed
            phase index.  (At the edge case $m=N$ of the $\eta_m$ construction
            above, $S_N=\{N\}$ gives $\ket{\phi^+}_{S_N}=\ket{+}_N$, so
            $\ket{\eta_N}=\ket{+}^{\otimes N}$ coincides with this case at
            $r_1=\cdots=r_N=0$.)  If $S\neq\varnothing$ but $S\neq\{1,\ldots,N\}$, the remaining
            computational-basis factors impose consistency constraints on the summation
            index $j$.  Whenever these constraints are consistent, the corresponding
            nonzero probabilities have equal weight $d^{-(|S|+1)}$; otherwise they vanish.
            In particular, when $|S|=N-1$, the depolarization state is the maximally mixed GHZ
            state.

\section{General finite phase span and simplex selection}
\label{app:general_finite_phase}

In this appendix, $N\ge2$ and $d\ge2$ are arbitrary, and
$\zeta=e^{2\pi i/m}$ with $m\ge3$.

\begin{proof}[Proof of the spanning assertion in Theorem~\ref{thm:general_finite_phase_simplex}]
Set $z_{s,i}=\zeta^{k_{s,i}}$, with $z_{s,0}=1$.
All computational
indices are reduced modulo $d$. Expanding the squared modulus in
Eq.~\eqref{eq:product_depolarization_probability} gives
\begin{equation}
p_{\lambda,\mathbf a}(\mathbf z)
=\frac1{d^{N+1}}\sum_{j,k\in\Z_d}
\om^{-\lambda(j-k)}
\prod_{s=1}^N z_{s,j+a_s}\overline{z_{s,k+a_s}}.
\label{eq:general_phase_character_expansion}
\end{equation}
The terms $j=k$ give the constant vector $p_{\mathrm{mix}}=d^{-N}$.
The remaining terms are nonconstant characters of the finite phase
group $(\Z_m)^{N(d-1)}$.

For $m\ge3$, a local character $z_i\overline{z_l}$ with $i\ne l$
uniquely identifies the ordered pair $(i,l)$: its exponent vector has
a $+1$ at $i$ and a $-1$ at $l$, omitting the fixed coordinate $0$.
The two signs are distinct modulo $m$. Consequently every product
character in Eq.~\eqref{eq:general_phase_character_expansion} with
$j\ne k$ identifies $(\mathbf a,j,k)$ uniquely. The first subsystem
determines $(j,k)$, and each later subsystem then determines $a_s$.

Let a real linear functional with coefficients $H_{\lambda,\mathbf a}$
be constant on all these phase probability vectors. Orthogonality of
characters on the finite phase group forces the coefficient of each
nonconstant character to vanish:
\begin{equation}
\sum_{\lambda\in\Z_d}H_{\lambda,\mathbf a}
\om^{-\lambda(j-k)}=0,
\qquad j\ne k.
\end{equation}
Every nonzero difference $j-k\in\Z_d$ occurs.
Thus all nonzero Fourier coefficients of
$H_{\lambda,\mathbf a}$ in $\lambda$ vanish, and $H$ is constant in
$\lambda$ on each fiber. These are exactly the functionals constant
on the slice in Eq.~\eqref{eq:general_uniform_fiber_mass}; there are
no additional affine constraints. This proves the claimed affine hull.

\end{proof}

\begin{proof}[Proof of the simplex assertion in Theorem~\ref{thm:general_finite_phase_simplex}]
The spanning assertion was established above.
\emph{Simplex selection.}
Uniform averaging over all phase choices in
Eq.~\eqref{eq:general_finite_phase_inputs} kills every nonconstant
character in Eq.~\eqref{eq:general_phase_character_expansion} and gives
$\rho_{\mathrm{mix}}$. By that spanning property, we can therefore
choose $(d-1)d^{N-1}$ phase states whose differences from
$\rho_{\mathrm{mix}}$ are linearly independent. They span the
zero-fiber-mass subspace. The $d^{N-1}$ fiber states supply
$d^{N-1}-1$ independent fiber-mass differences, and their average
is also $\rho_{\mathrm{mix}}$. Changing the reference for the selected
phase states to any one fiber state preserves the combined span,
since $\rho_{\mathrm{mix}}-\eta_{\mathbf 0}
=d^{1-N}\sum_{\mathbf a\neq\mathbf 0}(\eta_{\mathbf a}-\eta_{\mathbf 0})$
lies in the span of the fiber differences. The resulting $d^N$ vertices have
affine rank
\[
(d^{N-1}-1)+(d-1)d^{N-1}=d^N-1.
\]
All vertices are GHZ depolarizations of product states and are therefore
fully separable.
\end{proof}

\section{Three-qutrit graph realization and affine independence}
\label{app:three_qutrit_proof}

We return to $N=d=3$.

\begin{figure*}[t]
\centering
\begin{tikzpicture}[x=0.31cm,y=0.31cm,font=\small]
  \fill[black!3] (0,0) rectangle (18,18);
  \foreach \block in {0,...,8}{%
    \pgfmathtruncatemacro{\offset}{2*\block}
    \fill[phasezero!85] (\offset,16-\offset) rectangle ++(2,2);
  }
  \foreach \line in {0,...,18}{%
    \draw[white,line width=0.2pt] (\line,0) -- (\line,18);
    \draw[white,line width=0.2pt] (0,\line) -- (18,\line);
  }
  \foreach \line in {0,2,...,18}{%
    \draw[black!20,line width=0.35pt] (\line,0) -- (\line,18);
    \draw[black!20,line width=0.35pt] (0,\line) -- (18,\line);
  }
  \draw[black!45] (0,0) rectangle (18,18);
  \foreach \block/\slopes in
    {0/{(0,0)},1/{(1,2)},2/{(2,1)},3/{(0,2)},4/{(1,1)},
     5/{(2,0)},6/{(0,1)},7/{(1,0)},8/{(2,2)}}{%
    \node[font=\scriptsize,anchor=east] at (-0.3,17-2*\block) {$\slopes$};
    \node[font=\scriptsize,rotate=90,anchor=east]
      at (1+2*\block,-0.4) {$\slopes$};
  }
  \node[align=center] at (9,20.0)
    {Columns: centered phase states\\$c=0,1$ within each $(u,v)$ pair};
  \node[rotate=90,align=center] at (-5.0,9)
    {Rows: $M=(1,-u-v,u)$, $(2,u+v,-u)$\\within each $(u,v)$ pair};
  \node[anchor=west,align=left] at (20.5,13.5)
    {$\displaystyle\mathsf F^{\mathrm{ph}}=\begin{pmatrix}1&\om\\1&\om^2\end{pmatrix}$\\[8pt]
     $\det\mathsf F^{\mathrm{ph}}=-i\sqrt3\ne0$};
  \fill[phasezero!85] (20.5,7.8) rectangle ++(1,1);
  \node[anchor=west] at (22.0,8.3) {$|f_M[\mathbf g]|=1$};
  \fill[black!3,draw=black!25] (20.5,5.0) rectangle ++(1,1);
  \node[anchor=west] at (22.0,5.5) {$|f_M[\mathbf g]|=0$};
\end{tikzpicture}
\caption{Entrywise absolute values of the $18\times18$ phase Fourier
matrix.  Slope pairs follow the order in Eq.~\eqref{eq:tuple_graph_mapping};
each column pair has $c=0,1$, and each row pair has HW modes
$M=(1,-u-v,u)$ and $M=(2,u+v,-u)$, with all entries reduced modulo $3$.
The nine disjoint $2\times2$ blocks have unit-modulus entries.
Their complex phases, shown in $\mathsf F^{\mathrm{ph}}$, give a nonzero determinant
for every block. Together with the eight fiber directions, the phase rank $9\cdot2=18$
gives total affine dimension $26$.}
\label{fig:fourier_blocks}
\end{figure*}

\begin{proof}[Proof of Proposition~\ref{prop:three_qutrit_simplex}]
All polynomial coefficients and GHZ labels in this proof are modulo $3$.

\paragraph{Depolarized finite-phase states.}
Write $\braket{x}{u_{r,s}}=3^{-1/2}\om^{h_{r,s}(x)}$ with $h_{r,s}(0)=0$,
$h_{r,s}(1)=r$, $h_{r,s}(2)=s$. By Eq.~\eqref{eq:master_product_formula},
\begin{equation}
 p_{\lambda_1,a,b}^{(\mathbf r,\mathbf s)}
 =\frac1{81}\Bigl|\sum_{j=0}^2\om^{e_j(\lambda_1,a,b)}\Bigr|^2,
 \label{eq:discrete_phase_prob}
\end{equation}
with $e_j(\lambda_1,a,b)=-\lambda_1 j+h_{r_1,s_1}(j)+h_{r_2,s_2}(j\oplus a)+h_{r_3,s_3}(j\oplus b)$.
The function $h_{r,s}$ has the unique polynomial representation of degree at most two over $\Z_3$
\begin{equation}
 h_{r,s}(x)=(s-r)x+(2r-s)x^2,
 \qquad x\in\Z_3.
 \label{eq:phase_polynomial}
\end{equation}
The degree restriction is essential: adding a multiple of $x^3-x$
leaves all three values unchanged.
Indeed, its values at $x=0,1,2$ are $0,r,s$, respectively:
\begin{align}
 h_{r,s}(0)&=0, \notag\\
 h_{r,s}(1)&=(s-r)+(2r-s)=r, \notag\\
 h_{r,s}(2)&=2(s-r)+4(2r-s)=6r-2s\equiv s\pmod3.
\end{align}
For a finite-phase product input, set
\begin{equation}
 \gamma_\ell\coloneqq s_\ell-r_\ell,\qquad \xi_\ell\coloneqq 2r_\ell-s_\ell,
 \qquad \ell=1,2,3.
 \label{eq:phase_polynomial_coefficients}
\end{equation}
Here $\ell$ labels the subsystem. Then
$h_{r_\ell,s_\ell}(i)=\gamma_\ell i+\xi_\ell i^2$, and
\begin{align}
 h_{r_1,s_1}(j)&=\gamma_1j+\xi_1j^2, \notag\\
 h_{r_2,s_2}(j+a)&=\gamma_2(j+a)+\xi_2(j+a)^2
 \notag\\
 &=h_{r_2,s_2}(a)+\gamma_2j+\xi_2j^2+2\xi_2aj, \notag\\
 h_{r_3,s_3}(j+b)&=\gamma_3(j+b)+\xi_3(j+b)^2
 \notag\\
 &=h_{r_3,s_3}(b)+\gamma_3j+\xi_3j^2+2\xi_3bj.
\end{align}
Expanding the exponent in Eq.~\eqref{eq:discrete_phase_prob} gives
\begin{align}
 e_j(\lambda_1,a,b)
 &= (\xi_1+\xi_2+\xi_3)j^2 \notag\\
 &\quad+\bigl(-\lambda_1+\gamma_1+\gamma_2+\gamma_3\bigr)j \notag\\
 &\quad+(2\xi_2a+2\xi_3b)j \notag\\
 &\quad+h_{r_2,s_2}(a)+h_{r_3,s_3}(b).
 \label{eq:phase_exponent_expansion}
\end{align}
When $\xi_1+\xi_2+\xi_3=0$, define the graph labels by
\begin{equation}
 c\coloneqq\gamma_1+\gamma_2+\gamma_3,\qquad
 u\coloneqq 2\xi_2,\qquad v\coloneqq 2\xi_3.
 \label{eq:affine_graph_labels}
\end{equation}
The two terms in Eq.~\eqref{eq:phase_exponent_expansion} that are
independent of $j$ contribute only a phase of unit modulus.  Thus
\begin{align}
 p_{\lambda_1,a,b}^{(\mathbf r,\mathbf s)}
 &=\frac1{81}\left|\sum_{j=0}^2
          \om^{(c+ua+vb-\lambda_1)j}\right|^2 \notag\\
 &=\frac1{81}\left(3\,\delta_{\lambda_1,c+ua+vb}\right)^2 \notag\\
 &=\frac19\,\delta_{\lambda_1,c+ua+vb}.
 \label{eq:affine_graph_probability}
\end{align}
This is the probability distribution of Eq.~\eqref{eq:affine_graph_state}.
For the product in Eq.~\eqref{eq:graph_product_input}, the coefficients are
\[
\begin{gathered}
(\gamma_1,\gamma_2,\gamma_3)=(c,0,0),
\\
(\xi_1,\xi_2,\xi_3)=(u+v,-u,-v).
\end{gathered}
\]
Their quadratic coefficients sum to zero, and $2(-u)=u$, $2(-v)=v$
modulo $3$, proving the stated product realization.

When $\xi\coloneqq\xi_1+\xi_2+\xi_3\neq0$, completing the square ($2$ is invertible
modulo $3$) turns the sum in Eq.~\eqref{eq:discrete_phase_prob} into a quadratic
Gauss sum, $\bigl|\sum_{j\in\Z_3}\om^{\xi j^2+\beta j}\bigr|^2=3$ for every $\beta$.
Hence $p_{\lambda_1,a,b}=1/27$ for all $(\lambda_1,a,b)$, i.e., the depolarized
state is $\rho_{\mathrm{mix}}$. Since $(r_\ell,s_\ell)\mapsto(\gamma_\ell,\xi_\ell)$
is a bijection of $\Z_3^2$, exactly $3^5=243$ of the $729$ inputs have $\xi=0$;
they give each graph state $\rho^{\mathrm{ph}}_{c,u,v}$ nine times (the
$\gamma_\ell$ with $\gamma_1+\gamma_2+\gamma_3=c$), and the other $486$ give
$\rho_{\mathrm{mix}}$.

For the Fourier-block visualization, we order the slope pairs as
\begin{equation}
\begin{aligned}
(u,v)={}&(0,0),(1,2),(2,1),(0,2),(1,1),\\
        &(2,0),(0,1),(1,0),(2,2).
\end{aligned}
\label{eq:tuple_graph_mapping}
\end{equation}

\paragraph{Separating the fiber and phase directions.}
Identify GHZ-diagonal Hermitian operators with their real coefficient
vectors in the projectors $\Phi_{\lambda_1,a,b}$.
Define the eight fiber differences and the eighteen centered phase vectors by
\begin{equation}
\begin{gathered}
 \mathbf e_{a,b}\coloneqq\eta_{a,b}-\eta_{0,0},\quad (a,b)\ne(0,0),
 \\
 \mathbf g_{c,u,v}\coloneqq\rho^{\mathrm{ph}}_{c,u,v}-\rho_{\mathrm{mix}}.
\end{gathered}
 \label{eq:centered_directions}
\end{equation}
Replacing each original phase difference
$\rho^{\mathrm{ph}}_{c,u,v}-\eta_{0,0}$ by $\mathbf g_{c,u,v}$ preserves
the span of the 26 difference vectors, since
\begin{equation}
 \rho_{\mathrm{mix}}-\eta_{0,0}
 =\frac19\sum_{(a,b)\ne(0,0)}\mathbf e_{a,b}.
 \label{eq:reference_change}
\end{equation}
Therefore
\begin{equation}
 \rho^{\mathrm{ph}}_{c,u,v}-\eta_{0,0}
 =\mathbf g_{c,u,v}+\frac19\sum_{(a,b)\ne(0,0)}\mathbf e_{a,b}.
\end{equation}
Extend the fiber-mass map $q_0(a,b)$ linearly to these coefficient vectors,
\begin{equation}
 q_0(a',b')[\mathbf x]\coloneqq\sum_{\lambda_1=0}^2x(\lambda_1,a',b').
\end{equation}
Since
\begin{equation}
\begin{gathered}
 q_0(a',b')[\eta_{a,b}]=\delta_{a',a}\delta_{b',b},
 \\
 q_0(a',b')[\rho^{\mathrm{ph}}_{c,u,v}]
 =q_0(a',b')[\rho_{\mathrm{mix}}]=\frac19,
\end{gathered}
\end{equation}
for all $(a',b')\in\Z_3^2$,
\begin{equation}
\begin{gathered}
 q_0(a',b')[\mathbf e_{a,b}]
 =\delta_{a',a}\delta_{b',b}-\delta_{a',0}\delta_{b',0},
 \\
 q_0(a',b')[\mathbf g_{c,u,v}]=0.
\end{gathered}
 \label{eq:fiber_difference_coordinates}
\end{equation}
Restricting the fiber-mass coordinates to the eight non-reference pairs
$(a',b')\ne(0,0)$ makes the fiber block $I_8$.

\paragraph{Nine independent Fourier blocks.}
We use the Fourier coefficients $f_M$ defined in
Sec.~\ref{subsec:fourier} and the HW pairing in Eq.~\eqref{eq:ML_pairing}.
For $M=(\mu,\nu_1,\nu_2)$ and $\Lambda=(\lambda_1,a,b)$ in $\Z_3^3$,
the pairing is
\begin{equation}
 (M,\Lambda)=-\mu\lambda_1+\nu_2a-(\nu_1+\nu_2)b.
\end{equation}
We extend $f_M$ linearly to GHZ-diagonal Hermitian operators, identified
with their real coefficient vectors. For any such vector $\mathbf g$, set
\begin{align}
 f_M[\mathbf g]
 &\coloneqq\sum_{\Lambda\in\Z_3^3}g_\Lambda\,\om^{-(M,\Lambda)} \notag\\
 &=\sum_{\lambda_1,a,b\in\Z_3}
       g(\lambda_1,a,b)\,\om^{\mu\lambda_1-\nu_2a+(\nu_1+\nu_2)b}.
 \label{eq:centered_fourier_transform}
\end{align}
For a probability vector $\mathbf p$, $f_M[\mathbf p]$ is precisely the
coefficient in Eq.~\eqref{eq:ghz_fourier_coefficients}.
The centered phase vectors have
$f_{0,\nu_1,\nu_2}[\mathbf g_{c,u,v}]=0$ because their mass in every fiber is zero:
\begin{equation}
 f_{0,\nu_1,\nu_2}[\mathbf g]
 =\sum_{a,b\in\Z_3}q_0(a,b)[\mathbf g]\,
   \om^{-\nu_2a+(\nu_1+\nu_2)b}.
\end{equation}
The maximally mixed state has coefficients
\begin{align}
 f_M[\rho_{\mathrm{mix}}]
 &=\frac1{27}\sum_{\lambda_1}\om^{\mu\lambda_1}
 \sum_a\om^{-\nu_2a}\sum_b\om^{(\nu_1+\nu_2)b}
 \notag\\
 &=\delta_{\mu,0}\delta_{\nu_1,0}\delta_{\nu_2,0}.
\end{align}
Thus its contribution vanishes for $\mu=1,2$, and
Eq.~\eqref{eq:affine_graph_state} yields, for these values of $\mu$,
\begin{align}
 f_M[\mathbf g_{c,u,v}]
 &=\frac19\sum_{a,b\in\Z_3}
       \om^{\mu(c+ua+vb)-\nu_2a+(\nu_1+\nu_2)b} \notag\\
 &=\om^{\mu c}
   \left(\frac13\sum_a\om^{(\mu u-\nu_2)a}\right)
 \notag\\
 &\quad\times\left(\frac13\sum_b\om^{(\mu v+\nu_1+\nu_2)b}\right) \notag\\
 &=\om^{\mu c}\,\delta_{\nu_2,\mu u}\delta_{\nu_1,-\mu(u+v)}.
 \label{eq:phase_fourier_support}
\end{align}
For each slope pair $(u,v)$ the two vectors
$\mathbf g_{0,u,v},\mathbf g_{1,u,v}$ are therefore supported on exactly
the two Fourier coordinates $M=(1,-u-v,u)$ and $M=(2,u+v,-u)$:
\begin{align}
 f_{1,-u-v,u}[\mathbf g_{c,u,v}]&=\om^c, \notag\\
 f_{2,u+v,-u}[\mathbf g_{c,u,v}]&=\om^{2c}.
\end{align}
In that row order and with $c=0,1$ as columns, denote their Fourier
coordinate block by $\mathsf F^{\mathrm{ph}}_{u,v}$.  Its common value,
independent of $(u,v)$, is
\begin{equation}
\begin{gathered}
 \mathsf F^{\mathrm{ph}}_{u,v}=\mathsf F^{\mathrm{ph}}\coloneqq
 \begin{pmatrix}1&\om\\1&\om^2\end{pmatrix},
 \\
 \det\mathsf F^{\mathrm{ph}}=\om^2-\om=-i\sqrt3\ne0.
\end{gathered}
 \label{eq:phase_fourier_block}
\end{equation}
Distinct slope pairs have disjoint Fourier support, so the phase matrix is
$\bigoplus_{(u,v)\in\Z_3^2}\mathsf F^{\mathrm{ph}}$ and has rank $18$.
Invertibility over $\C$ implies independence of the original real vectors.

The fiber differences are uniform in $\lambda_1$, so all their Fourier
coefficients with $\mu=1,2$ vanish:
\begin{align}
 f_M[\mathbf e_{a,b}]
 &=\frac13\left(\sum_{\lambda_1=0}^2\om^{\mu\lambda_1}\right) \notag\\
 &\quad\times\left(\om^{-\nu_2a+(\nu_1+\nu_2)b}-1\right) \notag\\
 &=0,\qquad \mu=1,2.
\end{align}
Conversely, the centered phase vectors
have zero fiber masses.  In the eight non-reference fiber-mass coordinates
followed by the eighteen Fourier coordinates just specified, the transformed
difference matrix is thus
\begin{equation}
 \begin{pmatrix}
   I_8&0\\[2pt] 
   0&\displaystyle\bigoplus_{(u,v)\in\Z_3^2}\mathsf F^{\mathrm{ph}}
 \end{pmatrix}.
\label{eq:full_difference_blocks}
\end{equation}
Its determinant is
\begin{equation}
 \det I_8\cdot\left(\det\mathsf F^{\mathrm{ph}}\right)^9
 =\left(-i\sqrt3\right)^9\ne0.
\end{equation}
Every block is nonsingular, proving rank $8+9\cdot2=26$ and hence affine
independence of the 27 states.  Each fiber state is fully separable by
Eq.~\eqref{eq:eta_product_decomp}, and each phase state is the GHZ
depolarization of a pure product state.  Their convex hull is therefore
fully separable.
\end{proof}

Figure~\ref{fig:fourier_blocks} displays the nine Fourier blocks used in
the proof.

    \bibliography{GHZ_paper}
    \end{document}